# Magnetic-field dependent $V_B^-$ spin decoherence in hexagonal boron nitrides: A first-principles study

*Jaewook Lee[1,2], Hyeonsu Kim[1,2], Huijin Park[1,2], and Hosung Seo[1,2,3,4]**

J.Lee, H.Kim, H.Park, H.Seo

[1]SKKU Advanced Institute of Nanotechnology (SAINT), Sungkyunkwan University, Suwon, Republic of Korea

Email: seo.hosung@skku.edu

J.Lee, H.Kim, H.Park, H.Seo

[2]Department of Physics and Department of Energy Systems Research, Ajou University, Suwon, Republic of Korea

H.Seo

[3]Center for Quantum Information, Korea Institute of Science and Technology, Seoul, 02792, Republic of Korea

H.Seo

[4]Department of Quantum Information Engineering, Sungkyunkwan University, Suwon, 16419, Republic of Korea

**Funding:** This study is supported by the National Research Foundation (NRF) of Korea grant funded by the Korean government (MSIT) (No. 2023R1A2C1006270), by Creation of the Quantum Information Science R&D Ecosystem (Grant No. 2022M3H3A106307411) through the NRF of Korea funded by the Korea government (MSIT). This research was supported by the education and training program of the Quantum Information Research Support Center, funded through the National research foundation of Korea (NRF) by the Ministry of science and ICT(MSIT) of the Korean government (No.2021M3H3A103657313). This work was supported by Institute of Information & communications Technology Planning & Evaluation (IITP) grant funded by the Korea government (MSIT) (No.2022-0-01026). This work was supported by the National Supercomputing Center with supercomputing resources including technical support (ksc-2023-CRE-0321). This material is based upon work supported by, or in part by, the KIST institutional program (Project. No. 2E32971)

**Abstract**

The negatively charged boron vacancy ($V_B^-$) in h-BN is a spin-1 defect functioning as an optically addressable spin qubit in two-dimensional materials. A precise understanding of its spin decoherence is essential to advance it into a robust qubit platform. First-principles quantum many-body simulations are employed to investigate $V_B^-$ decoherence in dense nuclear spin baths of h-BN under magnetic fields from 0.01 to 3 T, considering isotopic variants h-$^{10}B^{14}N$, h-$^{11}B^{14}N$, h-$^{10}B^{15}N$, and h-$^{11}B^{15}N$. A transition boundary (TB) is observed where the dominant decoherence mechanism changes: below the TB, sub-microsecond decoherence is governed by independent nuclear spin dynamics, whereas above it, pairwise flip-flops dominate, extending $T_2$ to tens of microseconds. Analytical predictions place the TB at 0.502 T for h-$^{10}B^{14}N$ and 0.205 T for h-$^{11}B^{14}N$. The larger TB in h-$^{10}$BN results from the larger nuclear spin of $^{10}B$ ($I = 3$), which produces stronger nuclear modulation over a wider field range. The analytical approach also explains the magnetic-field-insensitive fast modulation observed below the TB. These findings clarify the role of dense nuclear spin baths with large nuclear spins ($I \geq 1$) in $V_B^-$ decoherence and provide design principles for isotopically engineered h-BN spin qubits.

## 1. Introduction

Quantum defects in hexagonal boron nitride (h-BN) have recently attracted significant attention for their potential as quantum sensing platforms, especially in imaging.[1–5] In 2016, a room-temperature single-photon emitter (SPE) was discovered in h-BN, which is attributed to point defects.[6] Subsequent research revealed a variety of bright SPEs in h-BN, spanning wavelengths from near-infrared to ultraviolet[7–9], and optically addressable spin defects[10–14]. In addition, the two-dimensional (2D) nature of the h-BN crystal allows for strong light-matter interactions, efficient photon extraction, and deterministic defect generation.[8,15] Furthermore, integration of h-BN quantum defects with various nano-photonic devices such as waveguides, cavities, and nano-trenches rapidly advances.[16–21] These properties have positioned quantum defects in h-BN as potential key components of integrated quantum photonics[2,4,22]. Along with the efforts at the materials level, the quantum applications of these defects are also being actively explored[23,24], including quantum simulation[25–27], quantum communication[28], quantum sensor[29,30].

One of the most notable quantum defects[22,31] in h-BN is the negatively charged boron vacancy ($V_B^-$) defect, identified by Abdi *et al.*[32], and Gottscholl *et al.*[33] as an optically addressable spin qubit. The defect has a spin-triplet ground state ($S = 1$), which exhibit a number of features reminiscent of the well-established diamond nitrogen-vacancy (NV) centers[34–37], including initialization, manipulation, and read out using a microwave and laser excitation at room temperature.[10] Notably, as a quantum defect in 2D host materials, the $V_B^-$ defect may offer opportunities not available in bulk systems.[3,29,38,39] In principle, the atomically thin structure of h-BN allows the $V_B^-$ defect to be positioned in close proximity to target samples, which is advantageous for high-resolution magnetic imaging and nanoscale sensing.[40–43] While the coherence properties and quantum efficiency of $V_B^-$ centers remain lower than those of NV centers, their integration in 2D platforms enables promising directions in device fabrication [44–47] and quantum nanodevice engineering [48]. These advantages led to immediate applications of the $V_B^-$ spin as an atomic-scale sensor of magnetism, temperature, and pressure in quantum materials.[29,39,40,49,50] However, the $V_B^-$ defect also faces inherent limitations, which include its low quantum efficiency (~0.03 %) and broad emission spectrum.[51,52] To address this challenge, several strategies have been explored to enhance the quantum yield of $V_B^-$ emission such as integrated plasmonic gap cavities[53,54].

Owing to these promising applications of $V_B^-$, numerous theoretical and experimental studies have been performed to analyze and control its spin coherent properties.[10,33,55] Initial efforts centered around characterizing the spin decoherence in h-BN. Theoretical studies[56–58] revealed that the spin decoherence is dominated by the dense spinful isotopes of boron and nitrogen, which are 19.9 % ($^{10}B$, $I = 3$), 80.1 % ($^{11}B$, $I = 3/2$), 99.6 % ($^{14}N$, $I = 1$), and 0.4 % ($^{15}N$, $I = 1/2$). Spin echo coherence time ($T_2$) of $V_B^-$ defect was also measured in several experimental studies, ranges from tens of nanoseconds to several microseconds.[40,42,43,58–63] Building upon these early studies, recent efforts have focused on enhancement and manipulation of the coherent property of the $V_B^-$ spin system, which comprises of the electron spin and the nearest-neighboring nuclear spins.[42,57,58] Notably, the synthesis of isotopically engineered h-BN samples was recently made possible. It was shown that $V_B^-$ defect in $^{10}$B- and $^{15}$N-enriched samples exhibited significantly less crowed spectra and dramatically narrower transitions in electron spin resonance spectroscopy than in natural h-BN, leading to improved quantum sensing performance.[43,64] In addition, large nuclear spin polarization[25,43,65] was demonstrated in an isotopically enriched h-BN sample, a key prerequisite for developing quantum registers and quantum simulators.

Despite the marked progress made by the previous studies, many fundamental properties of the spin decoherence process in the $V_B^-$ spin remain unknown. Indeed, a detailed understanding of the role of external magnetic field ($B_0$) and the surrounding nuclear spins in the spin decoherence is not established. In the magnetic field range up to 0.025 T, the Hahn-echo coherence of $V_B^-$ exhibits complex modulation and rapid decay[10,33,40,42,43,58,60–63]: $T_2$ = 82 ns (h-$^{nat}$BN)[61], 62 ns (h-$^{10}$BN)[58], 46 ns (h-$^{11}$BN)[58], 100 ns (h-$^{nat}$BN)[63], and 70 ns (h-$^{nat}$BN)[60] were reported in magnetic fields of 0 T, 0.015 T, 0.015 T, 0.02 T, and 0.025 T, respectively, which are consistent with a theoretical prediction $T_2$ = 115 ns (h-$^{10}$BN)[58] under 0.015 T. The reported data shows that the coherence does not saturate to its maximum value even above 0.025 T. Notably, the Hahn-echo coherence with a more gradual decay with enhanced $T_2$ were reported in a few tesla (T) range: 15.1 μs was experimentally measured[59] under 3.2 T and $T_2$ of 26 μs was theoretically predicted under 3 T[57]. The measured data indicates that the decoherence time-scale changes by more than two orders of magnitude between a few hundred Gauss and 3 T, but the transition behavior and its mechanism are unknown. This is also highly unusual in terms of the conventional picture of nuclear-spin-driven decoherence, which was mainly developed for diamond NV centers[66,67] and SiC spin qubits[68–70]. In those systems, the Hahn-echo coherence time saturates to its maximum value above a magnetic field strength of a few tens of Gauss, where all the nuclear spin flips other than pair-wise flip-flop transitions are suppressed due to the nuclear Zeeman effect. However, this picture clearly does not apply to the nuclear spin bath in h-BN, necessitating a systematic study of magnetic field-dependent decoherence mechanism of the $V_B^-$ spin.

In this study, we employ first-principles quantum many-body simulations to investigate the decoherence dynamics of the $V_B^-$ electron spin as a function of external magnetic field strength. Our results reveal a change in the dominant decoherence mechanism at a specific magnetic field strength, which we define as the transition boundary (TB). Using a general expression for electron-spin-echo-envelop-modulation (ESEEM), we demonstrate that below the TB, the $V_B^-$ spin echo is largely destroyed by strong nuclear modulation effects arising from the cancellation condition satisfied for large nuclear spins ($I \geq 1$) in the 100% dense nuclear spin bath of h-BN. Above the TB, these nuclear modulation effects are suppressed, and pair-wise flip-flop transitions in the nuclear spin bath become the primary decoherence source. We further find that the TB is influenced by the nuclear spin environment surrounding the $V_B^-$ defect, occurring at 0.502 T for h-$^{10}B^{14}N$ and 0.205 T for h-$^{11}B^{14}N$. Additionally, we identify key features intrinsic to the decoherence of $V_B^-$ in h-BN. These include the effect of ground-state level anti-crossing (GSLAC) and various echo modulation effects due to contact-hyperfine-induced independent nuclear dynamics and hyperfine-mediated nuclear spin flip-flop interactions. Our results help clarify the decoherence mechanism in h-BN, which exhibits increased complexity due to its denser and more diverse nuclear spin environment, resulting in spin dynamics that differ in character from those observed in diamond and SiC. This understanding paves the way for advancing $V_B^-$ qubits as robust platforms for quantum information science and technology in 2D materials.

## 2. System and model

To compute the decoherence dynamics of the $V_B^-$ spin, we employed a central spin model, in which the $V_B^-$ spin interacts with its surrounding nuclear spin environment under various magnetic field strengths. **Figure 1**a illustrates a schematic of the $V_B^-$ spin in multi-layer h-BN. An external magnetic field is applied along the crystal *c*-axis. We consider various magnetic field strengths in steps of 0.01 T from 0.03 to 0.5 T, in steps of 0.05 T from 0.5 to 1 T, and in steps of 0.1 T from 1 to 3 T. We also examine the effect different B and N isotopic compositions in h-BN on the spin decoherence. Specifically, we focus on the spin decoherence in h–$^{10}B^{14}N$ and h–$^{11}B^{14}N$[43,64]. To compute the hyperfine interaction due to the Fermi contact and anisotropic dipolar interactions, and the quadrupole interaction for nuclei with spin $I > ½$."[71], we used density functional theory (DFT) as implemented in the Vienna *ab-initio* simulation package (VASP)[72–74]. We used the Heyd–Scuseria–Ernzerhof exchange-correlation functional with the mixing parameter $\alpha = 0.32$ and the range-separation parameter $\omega = 0.2$[75,76] and DFT-3 method[77] to consider van der Waals (vdW) effect. The spin Hamiltonian parameters were computed using a four-layer h-BN supercell containing a single $V_B^-$ defect, which was verified to yield converged results through supercell size convergence test. Further details of the DFT methodology are provided in the Methods section.

We consider the Hahn-echo pulse sequence, which involves a π pulse applied between two free evolution time ($2\tau$). The decoherence of the $V_B^-$ spin is evaluated by using the off-diagonal element of the qubit's reduced density matrix. To calculate the effect of the large number of fluctuating nuclear spins, we employ the generalized cluster correlation expansion method (gCCE) at the gCCE2 level of theory, as implemented in the PyCCE code[78]. In our spin dynamics calculations, we aim to include as many h-BN layers as necessary to achieve numerically converged predictions of the $T_2$ time. The number of layers included is determined by a radius cutoff parameter, rBath, which defines the radius of the spin bath centered on the VB- defect. A value of rBath = 15 Å was found to yield converged coherence calculation, corresponding to nuclear spin contributions from nine h-BN layers around the defect, as shown in Supplementary Figure S1. In this study, we used rBath = 18 Å, which includes eleven layers of h-BN. Details on the spin Hamiltonian, theoretical methods, and the numerical convergence tests can be found in the Method section and Supporting Information Section S1 and S2.

### 3. Magnetic-field dependent spin decoherence

Figure 1b shows the computed coherence of the $V_B^-$ spin in h-$^{10}B^{14}N$ as a function of magnetic field from 0.01 to 3 T. The most important feature is the TB at a magnetic field strength of 0.502 T for h-$^{10}B^{14}N$, where the decoherence time-scale changes abruptly by two orders of magnitude. By fitting the coherence with a stretched exponential function ($\exp\left[-\left(\frac{2\tau}{T_2}\right)^n\right]$), we compute and plot the $T_2$ times in Figure 1c. It shows that the decoherence occurs within 0.35 μs below the TB, while above the TB, the decoherence time exceeds 70 μs. The fitting procedure and results corresponding to Figure 1c are shown in Figure S4 of the Supporting Information. Furthermore, we observe that the coherence function exhibits different modulation features depending on magnetic field regimes. As shown in Figure 1b, the coherence below the TB accompanies high-frequency modulation, which is invariant under changes of the magnetic field strength. In contrast, above the TB, the coherence function shown in Figure 1b retains a modulated feature with much lower frequency than that of the coherence below the TB. We also observe that modulation period varies as the magnetic field strength changes. By employing the Fourier transform (FT) analysis, we find that the primary modulation frequency is 47 MHz[42,61] for magnetic fields strength of 0.3 T below the TB, and 0.08 MHz and 0.07 MHz at $B_0$ = 2 T and 2.5 T, respectively, above the TB. The FT spectrum at $B_0$ = 0.3 T is shown in Figure S5 of the Supporting Information, and a detail analysis of the modulation frequency is discussed later. In addition, we observe a deviated coherence near $B_0$ = 1.42 T both in Figure 1b and 1d, which is due to an extra coherence modulation effect due to the $^{14}$N nuclear spins near the $V_B^-$. This feature and the related analysis are discussed in detail later in this paper.

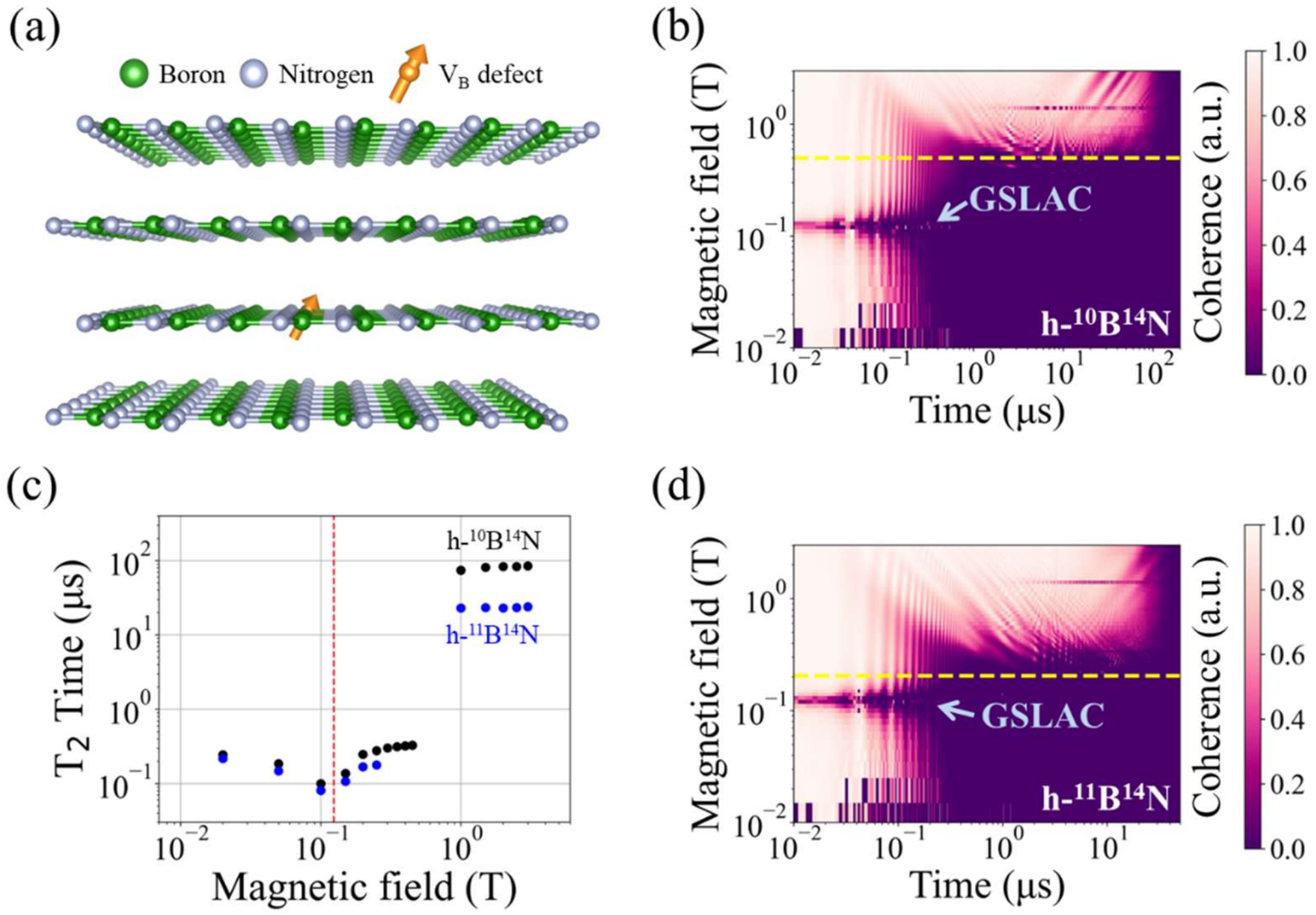

**Figure 1. Magnetic-field dependent decoherence of the $V_B^-$ spin. a)** Schematic of the $V_B^-$ spin and its surrounding bath in h-BN. **b,d)** Coherence function of the $V_B^-$ spin as a function of external magnetic field strength ranging from 0.01 to 3 T within a time scale up to 200 μs in h-$^{10}$B$^{14}$N (b), from 0.01 to 3 T within a time scale up to 50 μs in h-$^{11}$B$^{14}$N (d). The x-axis represents the free evolution time ($t = 2\tau$) in the Hahn-echo sequence, and the y-axis represents the strength of the magnetic field along the *c*-axis. The yellow dashed lines in (b) and (d) represent the transition boundaries (TBs), which are at 0.502 T and 0.205 T for h-$^{10}$B$^{14}$N and h-$^{11}$B$^{14}$N, respectively. **c)** $T_2$ time of the $V_B^-$ spin as a function of external magnetic field strength. The black and blue dots represent the $T_2$ time of the h-$^{10}$B$^{14}$N and h-$^{11}$B$^{14}$N, respectively. The red dashed line in the inserted figure indicates the ground-state level anti-crossing (GSLAC) of the $V_B^-$ defect.

Interestingly, we observe that the coherence undergoes extra decay under a magnetic field strength of 0.124 T as shown in Figure 1b,d. This phenomenon is attributed to the ground-state level anti-crossing (GSLAC) of the $V_B^-$ defect, where the electron spin states $|m_s\rangle$ = -1 and $|m_s\rangle$ = 0 in the ground-state cross, given the zero-field splitting (ZFS) of the $V_B^-$ defect to be 3.480 GHz and 50 MHz for D and E parameters[10], respectively. When the magnetic field is near the GSLAC, the electron spin experiences large population fluctuations that lead to a significant decrease in the coherence time.[70]

To understand the effect of the boron isotope on the decoherence, we compute the spin coherence in h-$^{11}$B$^{14}$N as shown in Figure 1b,d. Similar to the spin coherence in h-$^{10}$B$^{14}$N, we observe a TB but at a lower magnetic field strength around 0.2 T. Below the TB, the decoherence time is below 300 ns, while the decoherence timescale above the TB is increased by two orders of magnitude as shown in Figure 1c. We also find that the spin coherence in h-$^{11}$B$^{14}$N exhibits different modulation features depending on the magnetic field strength. The modulation frequency below the TB is 47 MHz and it is insensitive to magnetic field variation. Above the TB, we observe a coherence modulation at tens of μs timescale. We note, however, that the coherence modulation in h-$^{11}$B$^{14}$N above the TB is fainter than that of the coherence in h-$^{10}$B$^{14f}$N shown in Figure 1b. In what follows, we first discuss the decoherence below the TB more in detail, which is followed by analysis on the decoherence above the TB.

## 4. Decoherence below the transition boundary

To decipher the dominant decoherence mechanism at different magnetic field ranges, we compute the coherence in h-$^{10}$B$^{14}$N and h-$^{11}$B$^{14}$N at the gCCE1 level of theory in **Figure 2**a,b, respectively, which only capture the contribution from independent nuclear spin dynamics in the bath to the decoherence. By comparing these results to those in Figure 1, we find that gCCE1 reproduces the essential features of the decoherence below the TB, which include: (1) magnetic field-insensitive fast modulation with a frequency of 47 MHz, (2) a pronounced decoherence effect at 0.124 T (GSLAC), and (3) the TB at $B_0$ = 0.5 T and 0.2 T for h-$^{10}$B$^{14}$N and h-$^{11}$B$^{14}$N, respectively, below which the $T_2$ time is smaller than 0.4 μs. This comparison clearly shows that the decoherence below the TB is dominated by the independent nuclear spin dynamics.

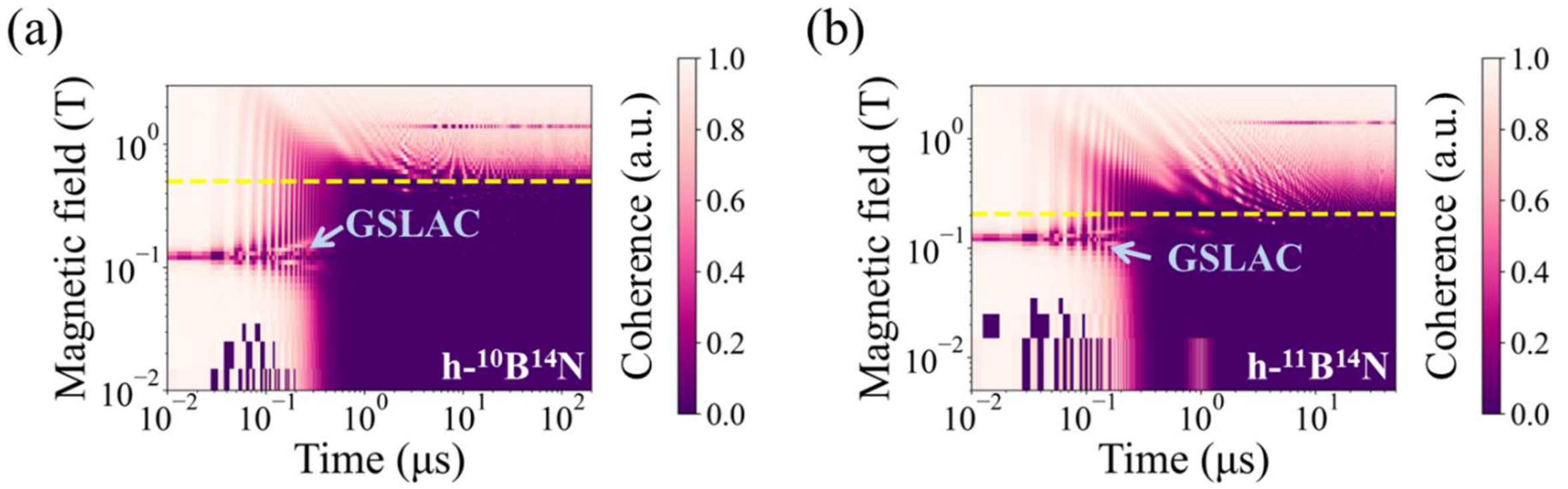

**Figure 2. $V_B^-$ decoherence in independent nuclear spin approximation. a,b)** Computed coherence function of the $V_B^-$ spin at the gCCE1 level of theory as a function of external magnetic field strength from 0.01 to 3 T in h-$^{10}B^{14}N$ within a time scale up to 200 μs in h-$^{10}B^{14}N$ (a), from 0.01 to 3 T in h-$^{11}B^{14}N$ within a time scale up to 50 μs in h-$^{11}B^{14}N$ (b). The yellow dashed line indicates the TB of the $V_B^-$ spin decoherence: 0.502 T for h-$^{10}B^{14}N$ and 0.205 T for h-$^{11}B^{14}N$. In (a) and (b), the observed extra coherence modulation near a magnetic field strength of 1.5 T is due to the $^{14}N$ nuclear spins near $V_B^-$ defect. The related analysis is discussed later.

It is surprising that the independent nuclear spin dynamics governs the decoherence under a large magnetic field strength up to thousands of Gauss. To better analyze its origin, we employ the general expression of electron spin echo envelop modulation (ESEEM)[79,80], which is the coherence between two qubit sublevels, i.e. $|\alpha\rangle$ and $|\beta\rangle$ of the $V_B^-$ spin affected by an arbitrary nuclear spin ($I$) located at a particular lattice site ($p$) under the secular approximation:

$$\mathcal{L}_p(2\tau) = V_0 + V_\alpha(2\tau) + V_\beta(2\tau) + V_+(2\tau) + V_-(2\tau) \quad (1)$$

, where $V_0$ is an unmodulated term that remains constant over time, while $V_\alpha$, $V_\beta$, $V_+$, and $V_-$ are modulated terms influenced by different nuclear frequencies. Their exact expressions are given in Supporting Information Section S3. The total coherence due to the entire nuclear spin bath is then given by the product of each nuclear spin contributions: $\mathcal{L}_{tot}(2\tau) = \prod_p \mathcal{L}_p(2\tau)$. It is worth noting that this is equivalent to the conventional CCE-1 (cCCE-1)[78] under secular approximation. Each modulation term, e.g. $V_\alpha$, consists of cosine modulations with various nuclear frequencies determined by nuclear Zeeman, hyperfine, and quadrupole interactions. And its modulation depth is determined by the so-called transition matrix ($\vec{M}$), whose matrix element $M_{ik}$ gives the electron paramagnetic resonance transition amplitude from the nuclear spin state $|i\rangle$ in the $|\alpha\rangle$ electron spin manifold to the nuclear spin state $|k\rangle$ in the $|\beta\rangle$ electron spin manifold. The value of $M_{ik}$ in each term in general ranges from 0 to 1, due to the diverse orientations of the eigenvectors of the nuclear spin Hamiltonian projected on the electron spin manifolds[79,80].

To see the behavior of each term in **Equation 1**, we select two random $^{11}$B nuclear spins near the $V_B^-$ spin in h-$^{11}$B$^{14}$N as shown in **Figure 3**a, and plot their $\mathcal{L}_p(2\tau)$, $V_\alpha + V_\beta$, and $V_+ + V_-$ in Figure 3b-e. Under a magnetic field strength of 0.05 T, the modulation in $\mathcal{L}_p(2\tau)$ is significant for both nuclear spins as shown in Figure 3b,d due to large modulation depth of $V_\alpha$, $V_\beta$, $V_+$, and $V_-$. Notably, the modulation frequencies of the two cases shown in Figure 3b,d significantly differ as the nuclear frequencies are determined not only by their Larmor frequency, but by the hyperfine interaction and the quadrupole interaction, which are position-dependent. On the other hand, if the magnetic field strength is increased to 0.5 T above the TB, the unmodulated $V_0$ is increased to 1, while the modulation depths for the $V_\alpha + V_\beta$, and $V_+ + V_-$ are reduced to zero as shown in Figure 3c,e.

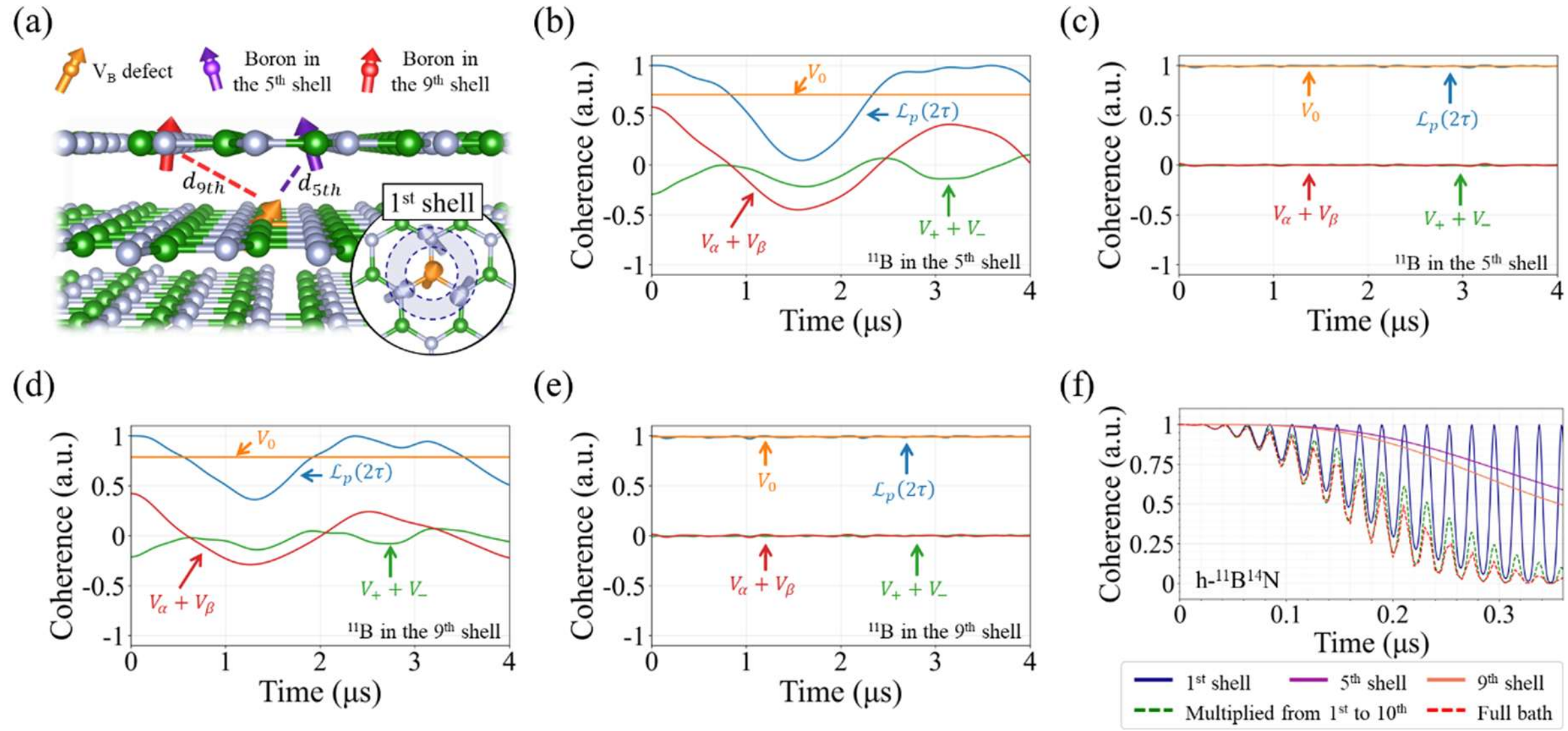

**Figure 3. Nuclear modulation effects in the $V_B^-$ decoherence. a)** Illustration of nuclear spins in different shells around $V_B^-$ in h-BN. The distance from $V_B^-$ to the 5$^{th}$ and 9$^{th}$ shells are denoted as $d_{5th}$ and $d_{9th}$, which are 2.9 Å and 3.8 Å. The inset figure shows nitrogen nuclear spins in the 1$^{st}$ shell, which are the nearest neighbors of the $V_B^-$ spin. **b-e)** Coherence function ($\mathcal{L}_p(2t)$) and individual terms ($V_0$, $V_\alpha + V_\beta$, $V_+ + V_-$) of the ESEEM equation for a single $^{11}$B nuclear spin. (b) and (c) correspond to the 5$^{th}$-shell nuclear spin at $B_0 = 0.05$ T and $B_0 = 0.5$ T, respectively, while (d) and (e) correspond to the 9$^{th}$-shell nuclear spin at $B_0 = 0.05$ T and $B_0 = 0.5$ T, respectively. **f)** Total coherence function constructed by multiplying the analytical ESEEM contributions from all nuclear spins in each shells (1$^{st}$ shell, 5$^{th}$ shell, etc.) and from the full nuclear spin bath (full bath) of h-$^{11}$B$^{14}$N at $B_0 = 0.05$ T. This represents the overall coherence at the first-order cluster level (CCE1) based on non-interacting single nuclear spins.

Figure 3f compares the total coherence function computed at the ESEEM equation under magnetic field strength of 0.05 T (the full bath data in the figure) to the coherence modulations due to the nuclear spins from various shells around the $V_B^-$ defect in h-$^{11}B^{14}N$. We find that the overall decoherence is captured in the product $\prod_p \mathcal{L}_p(2\tau)$. This means that the origin of the decoherence below the TB is the decay of the echo signal caused by the large number of different nuclear frequencies with deep modulations due to large nuclear spins ($I \geq 1$) of $^{11}B$ and $^{14}N$, and position-dependent hyperfine and quadrupole interactions.

The comparison shown in Figure 3f also reveals that the rapidly oscillating pattern in the coherence function below the TB arises due to the contribution from the three nearest neighboring $^{14}N$ nuclear spins in the 1$^{st}$ shell. In Supporting Information Section S4, we analytically derived the modulation frequencies due to the nearest neighboring $^{14}N$ nuclear spin, which are 1.3, 45.4, 46.7, and 47.9 MHz. The dominant frequency is 46.7 MHz, which is given as $\sqrt{(\omega_{14N} + A_{zz})^2 + \left(\frac{Q_{xx}-Q_{yy}}{2}\right)^2 + Q_{xy}^2}$, where $\omega_{14}$ is the Larmor frequency of the $^{14}N$, $A_{zz}$ is the hyperfine interaction, and $Q_{xx}$, $Q_{yy}$, and $Q_{xy}$ are the quadrupole tensor parameters. While $\omega_{14}$, $Q_{xx} - Q_{yy}$, and $Q_{xy}$ are smaller than 1 MHz in the magnetic field range from 0 T to 0.1 T, $A_{zz}$ is 47.6 MHz, which dominates the modulation frequency. Our result reveals the origin of the magnetic field-insensitive fast oscillation for $V_B^-$ spin both in h-$^{10}B^{14}N$ and h-$^{11}B^{14}N$, which is also in an excellent agreement with previous experimental observation[42,61]

It is well-established that nuclear modulation effects are greatly enhanced when the cancellation condition is met[80,81]. For nuclei with $I$ = 1/2 and 1[80,81], this condition occurs when the Zeeman and hyperfine interactions approximately cancel each other. Under such conditions, the energy splitting between nuclear spin sublevels ($\Delta$) becomes much smaller than the coupling strength ($\Omega$), resulting in strong mixing between sublevels. This strong mixing, in turn, leads to a pronounced increase in the depth of the modulation. However, for nuclei with spin $I$ > 1, such as $^{11}B$ and $^{10}B$, the analysis for the cancellation condition becomes more complex than a simple comparison of the Zeeman and hyperfine terms. In these cases, the quadrupole interaction and the full set of possible nuclear transitions must be taken into account in evaluating $\Delta$ and $\Omega$ (see Supplementary Information S5 and S6 for further details). **Figure 4**a illustrates the single-quantum and double-quantum transitions for a $^{11}B$ nuclear spin in h-BN. To determine whether the cancellation condition is satisfied in h-BN, we evaluate the ratio $|\Omega/\Delta|$ for each relevant transition and identify magnetic field regimes where this ratio exceeds unity.

Figure 4b and 4c present the computed ratio $|\Omega/\Delta|$ for all boron nuclear spins in h-$^{10}B^{14}N$ and h-$^{11}B^{14}N$, respectively. In both systems, this ratio exceeds unity across a broad range of magnetic field strengths, indicating the presence of strong nuclear modulation effects. By analyzing the energy levels of $^{10}B$ and $^{11}B$ nuclear spins as a function of magnetic field, we identify that the sharp peaks in $|\Omega/\Delta|$ correspond to level anti-crossings. These anti-crossings lead to strong sublevel mixing and enhanced ESEEM modulation, as also seen in a single-spin case of $^{11}B$ in Supplementary Figure S6. However, as the magnetic field strength increases further, the Zeeman interaction becomes the dominant term in the nuclear spin Hamiltonian. This results in a monotonic increase in sublevel splittings and the effective elimination of level anti-crossings. Consequently, the $|\Omega/\Delta|$ ratios for all nuclear spins eventually drops below unity, and the nuclear modulation effect is suppressed. This observation leads to a natural definition of the TB as the maximum magnetic field strength at which $|\Omega/\Delta| = 1$. Beyond this point, decoherence from independent nuclear spin dynamics is fully quenched. Our calculations yield the TBs to be 0.502 T for h-$^{10}B^{14}N$ and 0.205 T for h-$^{11}B^{14}N$. The difference in these values arises mainly due to the different gyromagnetic ratios of $^{10}B$ and $^{11}B$, which affect the magnetic field strength required to suppress the nuclear modulation (see Supplementary Figure S7).

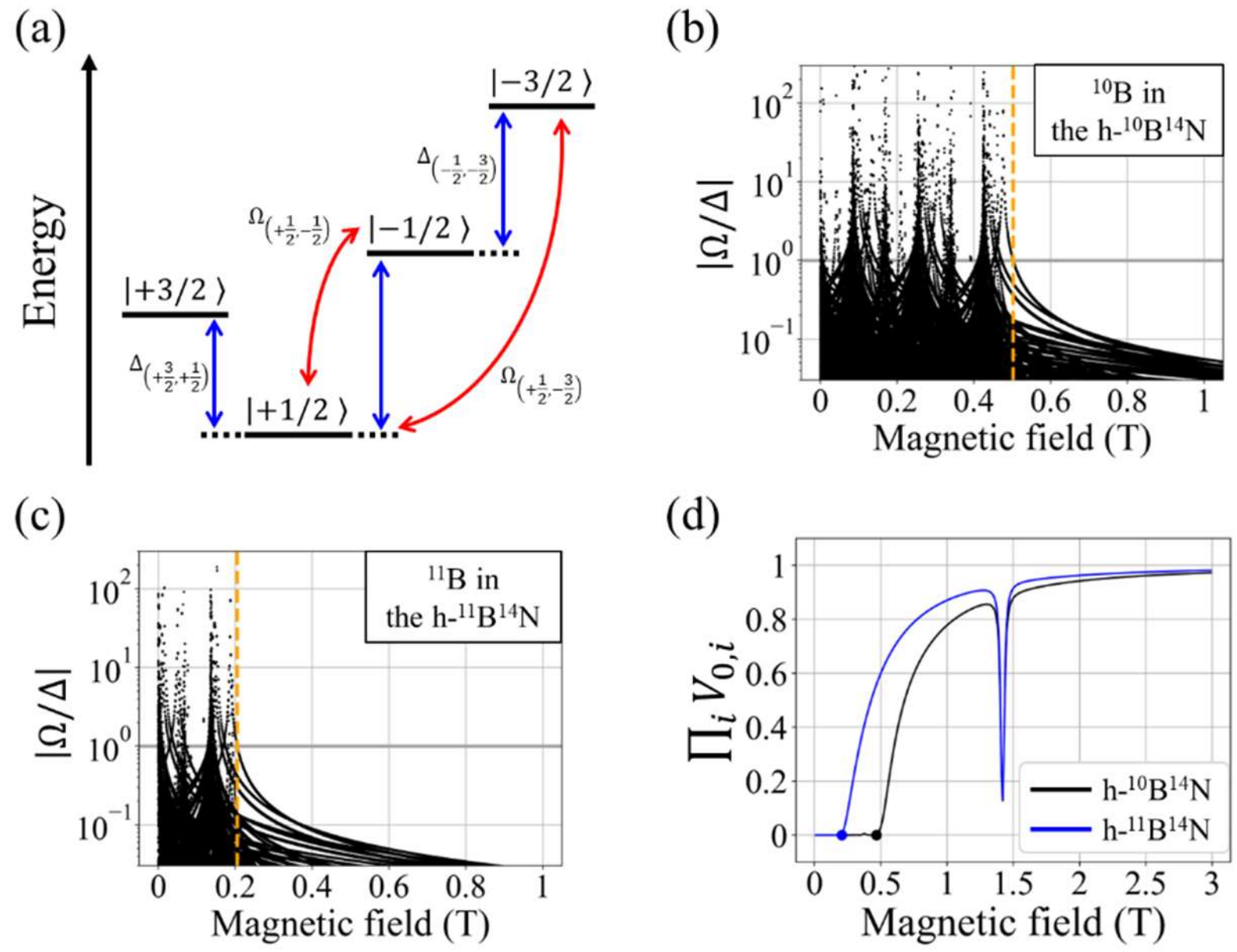

**Figure 4. Cancellation conditions in h-BN. a)** Schematic of spin sublevels for a single $^{11}$B nuclear spin subject to a magnetic field and the hyperfine field due to $V_B^-$. The blue and red arrows represent the energy difference ($\Delta_{(i,j)}$) and the coupling rate ($\Omega_{(i,j)}$) between two different states ($i$ and $j$), respectively. **b,c)** Ratio of the coupling rate to the energy difference ($|\Omega/\Delta|$) in the $V_B^-$'s $|+1\rangle$ electron spin manifold for all boron nuclear spins in the bath as a function of magnetic field strength: $^{10}$B nuclear spins in h-$^{10}$B$^{14}$N (b) and $^{11}$B nuclear spins in h-$^{11}$B$^{14}$N (c). The orange dashed line shows the largest magnetic field strength where $|\Omega/\Delta|$ is larger than 1, which we define as the TB: 0.502 T for h-$^{10}$B$^{14}$N and 0.205 T for h-$^{11}$B$^{14}$N. **d)** The total unmodulated part ($\prod_i V_{0,i}$) is plotted by multiplying all unmodulated terms from all nuclear spins in the bath (black line: h-$^{10}$B$^{14}$N, and blue line: h-$^{11}$B$^{14}$N). The computed x-intercepts (black and blue dots) are 0.485 T and 0.218 T for h-$^{10}$B$^{14}$N and h-$^{11}$B$^{14}$N, respectively, which closely estimate the position of the TB.

It is worth noting that the significance of the total nuclear modulation effect can also be collectively represented in the product of all unmodulated terms ($\prod_p V_{0,p}$) in Equation 1, where $p$ denotes individual nuclear spin in h-BN, which is shown in Figure 4d. It is because each $\mathcal{L}_p(2\tau)$ is bound to 1, and $V_{0,p}$ converges to 1 with increasing magnetic field strength as the other modulated terms would converge to 0. Below the TB, $\prod_p V_{0,p}$ is expected to be close to zero as each $V_{0,p}$ are less than 1. However, $\prod_p V_{0,p}$ would converge to 1 when the entire nuclear modulation effects are collectively suppressed above the TB. In Figure 4d, we find that $\prod_p V_{0,p}$ precisely exhibits this behavior, and that the x-intercepts for each data are 0.485 T for h-$^{10}$B$^{14}$N and 0.218 T for h-$^{11}$B$^{14}$N, closely estimating the TB.

## 5. Decoherence above the transition boundary

Above the TB, the nuclear modulation effect is suppressed, and the $T_2$ time is extended to tens of microseconds both in h-$^{10}B^{14}N$ and h-$^{11}B^{14}N$ as shown in Figure 1c. Thus, the decoherence is governed by spin correlations higher than the first order induced by the nuclear spin-spin interactions.[67] To determine the dominant order of spin correlations, we compare the coherence under $B_0$ = 3 T computed at the gCCE2 and gCCE3 levels of theory, which is shown in Figure S3a (Supporting Information). We find that the gCCE2 and gCCE3 results are almost the same in terms of its decay timescales. (However, their modulation patterns differ, and we discuss this in the following paragraphs more in detail.) Thus, we conclude that above the TB, the dominant decoherence mechanism is the pair-wise nuclear spin interactions. Notably, $T_2$ of the $V_B^-$ defect in h-$^{11}B^{14}N$ above the TB is much shorter than that in h-$^{10}B^{14}N$: e.g. at 1 T, the $T_2$ times in h-$^{11}B^{14}N$ and h-$^{10}B^{14}N$ are 23 μs and 74 μs, respectively. The smaller $T_2$ time in h-$^{11}B^{14}N$ is attributed to the small gyromagnetic ratio of the $^{10}B$ nuclear spin (2.875 rad $G^{-1}$ $ms^{-1}$), which is three times smaller than that of the $^{11}B$ nuclear spin (8.585 rad $G^{-1}$ $ms^{-1}$).[57]

In Figure 1b,d, we observed that the coherence above the TB shows a significant modulation on a timescale of tens of microseconds, which exhibits a dependence on the magnetic field strength. To shed light on its origin, in **Figure 5**a, we compare the coherence function of the $V_B^-$ spin in h-$^{10}B^{14}N$ under a large magnetic field strength of 3 T, computed using the gCCE2 method, conventional CCE2 (cCCE2)[78], and cCCE2 including hyperfine-mediated nuclear spin-spin interactions, which arise as second-order perturbations.[82] While we observe no modulation in the cCCE2 result, we find that the modulation is well reproduced in the cCCE2+perturbation method, indicating that the coherence modulation found in the gCCE2 coherence is derived from nuclear spin-spin interaction mediated by the central qubit spin[78,83–85].

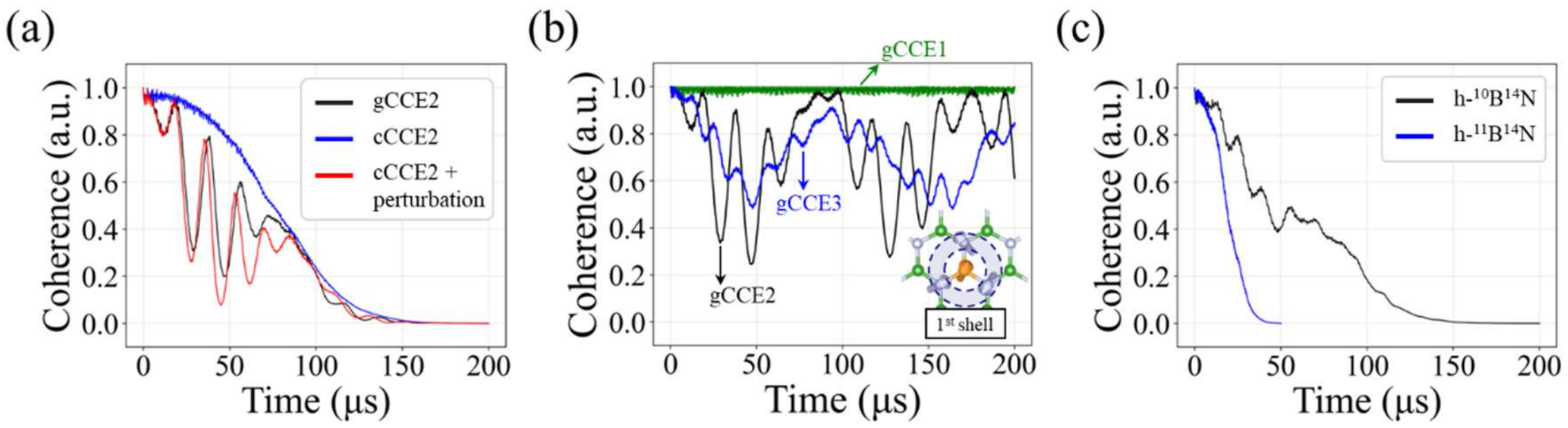

**Figure 5 Coherence modulation at a large magnetic field above the TB. (a)** Coherence function of the $V_B^-$ spin in h-$^{10}B^{14}N$ under a large magnetic field strength ($B_0$ = 3 T) computed using various simulation methods: gCCE2 (black), conventional CCE2 (cCCE2, blue), and cCCE2 with a hyperfine-mediated nuclear-nuclear perturbative interaction (red). **(b)** Coherence function of the $V_B^-$ spin computed by considering only the three $^{14}N$ nuclear spins in the 1st shell with different cluster order of gCCE method. The gCCE3 calculation is the same as the exact diagonalization. **(c)** Coherence function of the $V_B^-$ spin computed by combining the exact diagonalization of the 4-spin cluster (the $V_B^-$ spin and the three nearest-neighbor $^{14}N$ nuclear spins) with the gCCE2 calculation for the rest of the nuclear spin bath in h-$^{10}B^{14}N$ and h-$^{11}B^{14}N$ at $B_0$ = 3 T.

To further narrow down the origin of the coherence modulation above the TB, we consider a minimal 4-spin model, comprising of the $V_B^-$ spin and the three nearest-neighbor $^{14}N$ nuclear spins. We find that the same modulation is obtained in the gCCE2 result for this model (but with no decay). This comparison shows that the hyperfine-mediated spin-spin interaction among the three nearest-neighbor $^{14}N$ nuclear spins is the origin of the modulation. We find, however, that the modulation frequency and the depth are different between the gCCE2 and gCCE3 results for this 4-spin model. Considering that the gCCE3 result corresponds to the exact solution for the 4-spin model, the modulation found in the gCCE3 result is a more precise prediction for the modulation feature above the TB than the gCCE2 prediction.

To incorporate the exact modulation feature to the decoherence of the $V_B^-$ spin above the TB, we combine the exact diagonalization of the 4-spin cluster (the $V_B^-$ spin and the 3 nearest-neighbor $^{14}N$ nuclear spins) and the gCCE2 calculation for the rest of the nuclear spin bath. Computed coherence functions are shown in Figure 5c. We find that the $T_2$ times are the same as those of the gCCE2 prediction shown in Figure 1, while the modulation depth is reduced for both h-$^{10}B^{14}N$ and h-$^{11}B^{14}N$. We find, however, that significant modulation is still visible in h-$^{10}B^{14}N$, which may be detectable in experiment.

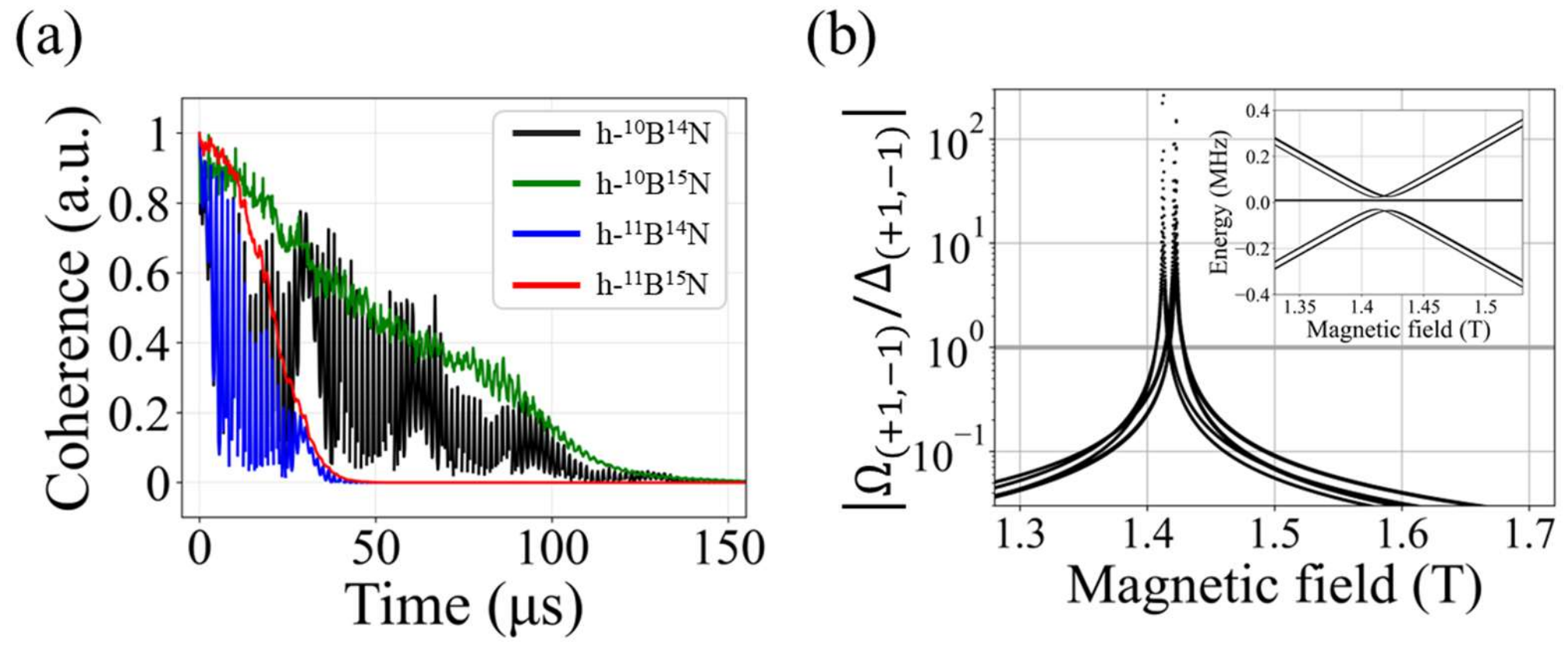

**Figure 6. Extra modulation of the $V_B^-$ coherence near $B_0$ = 1.42 T. a)** Computed coherence of the $V_B^-$ defect spin under a magnetic field strength of 1.42 T in various isotopically purified h-BN. The black and blue solid lines represent data for $^{10}B^{14}N$ and $^{11}B^{14}N$, respectively, while the green and red solid lines represent data for $^{10}B^{15}N$ and $^{11}B^{15}N$, respectively. **b)** The ratio of the coupling rate to the energy level difference between nuclear spin states +1 and -1 ($\Omega_{(+1,-1)}/\Delta_{(+1,-1)}$) of the $^{14}N$ nuclear spins in the 6th shell, subject to the hyperfine field of the $V_B^-$ defect spin in it spin sublevel of +1. The inset figure shows the energy level diagram of the $^{14}N$ nuclear spins in the 6th shell.

Interestingly, we observe an extra coherence modulation emerging at a magnetic field strength of 1.42 T as shown in **Figure 6**a. This feature is also visible in Figure 1b, and 1d for $^{10}B^{14}N$ and $^{11}B^{14}N$, respectively. We find that this modulation originates from the cancellation condition being satisfied for $^{14}N$ nuclear spins located in the 6th shell from the $V_B^-$ defect (N3 in Supplementary Figure S9). Figure 6b shows that the ratio Ω/Δ between the +1 and -1 states of the $^{14}N$ nuclear spin in the $V_B^-$'s $m_s$ = +1 electron spin manifold, which exhibits a sharp peak around 1.42 T. According to our DFT calculations, the secular hyperfine coupling ($A_{zz}$) for the 6th-shell $^{14}N$ nuclear spins is 4.35 MHz, due to a large Fermi-contact contribution computed to be 5.21 MHz (see Supplementary Table S1), which is also similar to a previous result[86]. Together with $Q_{xx} \approx 0.03$ MHz and $Q_{yy} \approx -0.02$ MHz (see Supplementary Table S2), Ω/Δ becomes much larger than 1 at 1.42 T. We also note that no other nuclear spins in h-BN satisfy this condition at nearby magnetic field strengths, leading to the sharp feature at 1.42 T without any gradual transition behavior across the magnetic field strength.

Additionally, we find that the fast modulation frequency observed in Figure 6a at $B_0$ = 1.42 T is 4.37 MHz, which is dominated by the contact hyperfine interaction and the Larmore frequency (see Equation S37 provided Supporting Information Section S4). We also remark that the fast modulation at $B_0$ = 1.42 T is not found in the coherence of the $V_B^-$ spin in h-$^{10}B^{15}N$ and h-$^{11}B^{15}N$, since the cancellation condition is not satisfied for the nuclear spin 1/2 of $^{15}N$ in the 6$^{th}$ shell (see Figure S8 in Supporting Information). Our results show that a magnetic field strength of 1.42 T is a special one for $V_B^-$ in h-$B^{14}N$, where two different modulation effects are simultaneously active: a fast one due to single nuclear spin dynamics under the cancellation condition and a slower one due to correlated multi-nuclear spin dynamics.

**6. Conclusion**

In summary, we investigated the decoherence of the $V_B^-$ spin in h-$^{10}B^{14}N$ and h-$^{11}B^{14}N$ as a function of magnetic field strength. We find that the $T_2$ time stays below 350 nanoseconds under magnetic field strengths of hundreds of Gauss. Surprisingly, we find an abrupt change of the decoherence timescale by two orders of magnitude above a specific magnetic field, which we refer to as the TB, which are 0.502 T and 0.205 T for the $V_B^-$ spin in h-$^{10}B^{14}N$ and h-$^{11}B^{14}N$, respectively. Furthermore, we show that the TB is where the decoherence mechanism changes: below the TB, the decoherence is dominated by independent nuclear spin dynamics, whereas above the TB, pair-wise flip-flop transitions become the primary source of decoherence. We identified that below the TB, the spin echo of the $V_B^-$ spin is largely destroyed due to significant nuclear modulation effects due to the cancellation condition satisfied for a wide range of magnetic field strength. The TB in $^{10}B^{14}N$ is larger than in $^{11}B^{14}N$ due to the larger nuclear spin of $^{10}B$ than that of $^{11}B$ and the smaller gyromagnetic ratio of the $^{10}B$ nuclear spin than that of the $^{11}B$ nuclear spin. Our results highlight the significant role of both the magnetic field and large nuclear spins ($I \geq 1$) in determining the decoherence properties of the $V_B^-$ defect.

We also find intriguing coherence modulation in the $V_B^-$ spin coherence, which changes its characteristic depending on the magnetic field. Below the TB, we find a magnetic field-insensitive fast modulation with a frequency of 47 MHz regardless of boron isotopes in h-$B^{14}N$, which was also observed in recent experiment[42,61]. We analytically identified that this fast modulation originates from the ESEEM due to the nearest neighboring $^{14}N$ nuclear spins, and its frequency is dominated by the contact hyperfine interaction. Above the TB, we find much slower coherence modulation in h-$^{10}B^{14}N$ than that below the TB. In addition, the modulation frequency in this case is dependent on the magnetic field strength. Notably, we find that the coherence modulation above the TB is not due to conventional ESEEM originating from single nuclear spin dynamics, but due to correlated multi-nuclear-spin dynamics induced by the hyperfine-mediated nuclear spin-spin interactions between the three nearest neighboring $^{14}N$ nuclear spins.

In our study, we mainly discussed the decoherence in h-B$^{14}$N. We note, however, our results and analysis remain valid for h-B$^{15}$N. In Supplementary Figure S8, we show the computed coherence of the $V_B^-$ spin in h-$^{11}$B$^{15}$N. By comparing these results and the results shown in Figure 1, we find that the TB and the overall coherence time remain essentially unchanged in h-B$^{15}$N. This is consistent with our analysis, as the decoherence timescale in h-BN is primarily governed by boron nuclear spins, rather than nitrogen. Therefore, enriching nitrogen from $^{14}$N to $^{15}$N does not significantly affect the dominant decoherence features in these systems. However, we find that the fast-oscillating modulations observed at low magnetic fields (below the TB) and the abrupt change in the decoherence profile at $B_0$ = 1.42 T in h-$^{11}$B$^{14}$N are absent in the $^{15}$N-enriched sample. These differences are also well explained within our framework: the modulation features in h-$^{11}$B$^{14}$N stem from $^{14}$N nuclear spins, specifically those located in the 1$^{st}$ and 6$^{th}$ shells relative to the $V_B^-$ defect. Since $^{15}$N has a different nuclear spin number and hyperfine structure, such features do not appear in the $^{15}$N-enriched system.

While our study provides important insights into the decoherence behavior of the $V_B^-$ spin in h-BN, we acknowledge several limitations in our approach. The coherence functions were calculated using the gCCE method, with cluster order truncated at three due to significant computational cost. In addition, the structural model employed assumes an ideal, undistorted multilayer h-BN lattice containing only boron and nitrogen nuclear spins. Potential decoherence sources such as strain, structural defects, or paramagnetic impurities were not considered. These simplifications may contribute to discrepancies between our theoretical predictions and experimental observations. Furthermore, although our calculations model isolated $V_B^-$ defect, real experiments typically involve ensembles of defects in non-ideal environments. In such setting, local variations in structure, impurity distributions, or defect clustering may broaden or obscure the nuclear modulation features predicted in this study. Therefore, while the modulation patterns we report are intrinsic to the spin–bath interaction for isolated $V_B^-$ defects, their experimental visibility may be reduced. To bridge this potential gap between theory and experiment, future studies could incorporate more realistic structural models and explicitly include other sources of decoherence beyond the nuclear spin bath.

To conclude, our study provides a unified framework for interpreting spin decoherence in h-BN over a wide range of magnetic field strengths. Particularly, our study unravels the distinctive role of dense spin baths with large nuclear spins ($I \geq 1$) in the qubit decoherence, paving the way to analyze and predict decoherence of spin qubits in various materials with similar bath properties.

We were informed during the preparation of the manuscript by a similar parallel work of complementing results.[87]

## 7. Method

### 7.1. Cluster correlation expansion method

Spin qubit decoherence results from its entanglement with the surrounding environment. In our study, we consider nuclear spins within the h-BN lattice serve as the main environmental degrees of freedom. These nuclear spins are coupled to each other through magnetic dipolar interactions and to the qubit via hyperfine interactions. The full spin Hamiltonian for the combined qubit and bath system is described in Supporting Information Section S1. To simulate the decoherence dynamics of the $V_B^-$ defect in this nuclear spin bath, we employ the generalized cluster-correlation expansion method (gCCE) as implemented in the PyCCE code.[78] For the gCCE calculations in the multilayer h-BN, we set a cluster order of 2, a bath size of 18 Å, and a maximum dipolar coupling radius between nuclear spins of 8 Å. (Details in the Supporting Information Section S2)

### 7.2. Density functional theory

We compute the electronic structure, quadrupole interaction, and hyperfine interaction of $V_B^-$ by using density functional theory as implemented in the Vienna ab initio simulation package (VASP).[72–74] We consider multi-layer h-BN with the AA' stacking.[75] We use a cutoff energy of 600 eV for the plane-wave expansion along with 8×8×8 k-point sampling, and a projector-augmented wave (PAW) pseudopotentials as implemented in VASP[88]. We employed Heyd–Scuseria–Ernzerhof exchange-correlation functional (HSE functional with $\alpha = 0.32$ and $\omega = 0.2$)[75,76], and VDW DFT-3 method[77] to consider van der Waals (vdW) interaction. To compute the $V_B^-$ defect in the h-BN, we construct a (7×4×2) orthogonal supercell. For the supercell calculation, we used a single $\Gamma$ point for the k-point sampling. We perform structural relaxation until the forces on each atom are less than 0.01 eV/ Å.

**Acknowledgements**

This study is supported by the National Research Foundation (NRF) of Korea grant funded by the Korean government (MSIT) (No. 2023R1A2C1006270), by Creation of the Quantum Information Science R&D Ecosystem (Grant No. 2022M3H3A106307411) through the NRF of Korea funded by the Korea government (MSIT). This research was supported by the education and training program of the Quantum Information Research Support Center, funded through the National research foundation of Korea (NRF) by the Ministry of science and ICT(MSIT) of the Korean government (No.2021M3H3A103657313). This work was supported by Institute of Information & communications Technology Planning & Evaluation (IITP) grant funded by the Korea government (MSIT) (No.2022-0-01026). This work was supported by the National Supercomputing Center with supercomputing resources including technical support (ksc-2023-CRE-0321). This material is based upon work supported by, or in part by, the KIST institutional program (Project. No. 2E32971)

**Data Availability Statement**

The data that support the findings of this study are available upon reasonable request to the corresponding author

**Supporting Information for: Magnetic-field dependent $V_B^-$ spin decoherence in hexagonal boron nitrides: A first-principles study**

*Jaewook Lee[1,2], Hyeonsu Kim[1,2], Huijin Park[1,2], and Hosung Seo[1,2,3,4]**

J.Lee, H.Kim, H.Park, H.Seo
[1]SKKU Advanced Institute of Nanotechnology (SAINT), Sungkyunkwan University, Suwon, Republic of Korea
Email: seo.hosung@skku.edu

J.Lee, H.Kim, H.Park, H.Seo
[2]Department of Physics and Department of Energy Systems Research, Ajou University, Suwon, Republic of Korea

H.Seo
[3]Center for Quantum Information, Korea Institute of Science and Technology, Seoul, 02792, Republic of Korea

H.Seo
[4]Department of Quantum Information Engineering, Sungkyunkwan University, Suwon, 16419, Republic of Korea

## S1. Generalized CCE (gCCE) method and Spin Hamiltonian

The decoherence of the qubit is determined by the decay of the off-diagonal elements in a qubit's density matrix. When the qubit is initially prepared in the state $|\psi\rangle = \frac{1}{\sqrt{2}}\left(|0\rangle + e^{i\phi}|1\rangle\right)$, the coherence function ($\mathcal{L}(2\tau)$) quantifies the loss of the relative phase between the $|0\rangle$ and $|1\rangle$ qubit levels:

$$\mathcal{L}(2\tau) = \frac{\langle 0|\rho_S(2\tau)|1\rangle}{\langle 0|\rho_S(0)|1\rangle} \tag{S1}$$

, where $\rho_S(2\tau)$ represents the density matrix of the central spin, and $|0\rangle$ and $|1\rangle$ are the qubit levels. However, in the presence of a bath of interacting nuclear spins, one needs to compute the dynamics of the reduced density matrix of the qubit, which becomes computationally infeasible task. To numerically solve this problem, we employ the cluster correlation expansion (CCE) method[1–3]. The core concept of the CCE method is that the coherence function is factorized into a set of irreducible contributions from the bath spin clusters. The CCE method is described as follows:

$$\mathcal{L}(2\tau) = \prod_i \tilde{\mathcal{L}}_{\{i\}}(2\tau) \prod_{i,j} \tilde{\mathcal{L}}_{\{i,j\}}(2\tau) \prod_{i,j,k} \tilde{\mathcal{L}}_{\{i,j,k\}}(2\tau) \ldots = \prod_C \tilde{\mathcal{L}}_C(2\tau) \tag{S2}$$

, where $\tilde{\mathcal{L}}_C(2\tau)$ is an authentic cluster correlation due to a cluster *C*, defined as follows:

$$\tilde{\mathcal{L}}_C(2\tau) = \frac{\mathcal{L}_C(2\tau)}{\prod_{C'} \tilde{\mathcal{L}}_{C'}(2\tau)} \tag{S3}$$

, where $\mathcal{L}_C(2\tau)$ is the coherence function of the qubit interacting only with the bath spins in a given cluster $C$, and $\tilde{\mathcal{L}}_{C'}(2\tau)$ represents the contribution of a subcluster $C'$ of $C$. There are two CCE methods developed in the literature: generalized CCE (gCCE) method[3] and conventional CCE (cCCE) method[3]. While the CCE method adopts the secular approximation for the hyperfine interaction, the gCCE method employs the full hyperfine interaction by including the central spin degrees of freedom into each cluster. The time evolution of $\mathcal{L}_C(2\tau)$ is described as follows:

$$\mathcal{L}_C(2\tau) = \left\langle 0 \middle| U_C(2\tau)\rho_{C+S}(2\tau)U_C^{\dagger}(2\tau) \middle| 1 \right\rangle \tag{S4}$$

, where $\rho_{C+S} = \rho_C \otimes \rho_S$ represents an initial density matrix of the combined bath spins ($\rho_C$) and central spin ($\rho_S$). For the Hahn-echo sequence involving a π rotation about the *x*-axis, the propagator is given by,

$$U_C(2\tau) = e^{-\frac{i}{\hbar}H_C\tau} e^{-\frac{i}{\hbar}\sigma_x\frac{\pi}{2}} e^{-\frac{i}{\hbar}H_C\tau} \tag{S5}$$

, where $\sigma_x$ is the Pauli matrix. And the full spin Hamiltonian ($H_C$) of bath spin clusters given as follows:

$$H_C = H_e + H_{ba} \quad + H_{e-bath} \tag{S6}$$

In Equation S6, the electron spin Hamiltonian ($H_e$) includes both the Zeeman effect and zero-field interaction:

$$H_e = -\gamma_e \vec{\boldsymbol{B}}_0 \cdot \vec{\boldsymbol{S}} + \vec{\boldsymbol{S}} \cdot \overleftrightarrow{\mathbf{D}} \cdot \vec{\boldsymbol{S}} \tag{S7}$$

, where $\gamma_e$ is the gyromagnetic ratio of the $V_B^-$ electron spin, $\overleftrightarrow{\mathbf{D}}$ is the zero-field splitting (ZFS) tensor, $\vec{\boldsymbol{S}}$ is the electron spin operator, and $\vec{\boldsymbol{B}}_0$ is the external magnetic field. The nuclear spin Hamiltonian ($H_{bath}$) consists of the Zeeman effect, quadrupole interactions, and dipole-dipole interactions between the nuclear spins in a bath cluster:

$$H_{bath} = -\vec{B}_0 \cdot \sum_i \gamma_{n_i} \vec{\boldsymbol{I}}_i + H_{n-n} + \sum_i \vec{\boldsymbol{I}}_i \cdot \overleftrightarrow{\mathbf{Q}}_i \cdot \vec{\boldsymbol{I}}_j \tag{S8}$$

$$H_{n-n} = \frac{\mu_0}{4\pi} \sum_{\{i,j\}} \gamma_{n_i} \gamma_{n_j} \left[ \frac{\vec{\boldsymbol{I}}_i \cdot \vec{\boldsymbol{I}}_j}{r_{ij}^3} - \frac{3(\vec{\boldsymbol{I}}_i \cdot \vec{\boldsymbol{r}}_{ij})(\vec{\boldsymbol{I}}_j \cdot \vec{\boldsymbol{r}}_{ij})}{r_{ij}^5} \right] \tag{S9}$$

, where $\gamma_{n_i}$ and $\vec{\boldsymbol{I}}_i$ are the gyromagnetic ratio and the spin operator of the *i*-th nuclear spin. $\vec{\boldsymbol{r}}_{ij}$ and $r_{ij}$ is the vector and distance between the *i*-th and *j*-th nuclear spins. $\overleftrightarrow{\mathbf{Q}}_i$ is a quadrupole interaction tensor of the *i*-th nuclear spin[4]. Finally, the interaction between the electron spin and nuclear spins ($H_{e-bat}$ ) is given by hyperfine interaction:

$$H_{e-bat} \quad = \vec{S} \cdot \sum_i \overleftrightarrow{\mathbf{A}}_i \cdot \vec{I}_i \tag{S10}$$

, where $\overleftrightarrow{\mathbf{A}}_i$ is the hyperfine tensor of the *i*-th nuclear spin.

## S2. Convergence test of the gCCE calculation

We performed numerical convergence tests to determine the appropriate termination point for the gCCE expansion, the bath size (rBath), which is defined as the distance between the central spin and the outermost nuclear spin, and the maximum dipolar coupling radius between nuclear spins (rDip). Figure S1,2 presents the results of the convergence tests under different magnetic field strengths (0.03 T, 0.3 T, and 3 T) and bath configurations (h–$^{10}B^{14}N$ and h–$^{11}B^{14}N$). We find that our gCCE calculations are converged with a bath size of 18 Å and a dipolar coupling radius of 8 Å.

Figure S3a,b shows the $V_B^-$ decoherence in h-$^{11}B^{14}N$ computed at the gCCE1, gCCE2, and gCCE3 levels of theory under magnetic fields of 3 T and 0.03 T, respectively. We used a bath radius (rBath) of 15 Å and a dipolar interaction distance (rDip) of 6 Å to reduce the computational cost for these calculations. At 3 T, which is above the transition boundary (TB), the gCCE2 and gCCE3 calculations yield similar decoherence, while the gCCE1 result does not show any decoherence. Notably, the gCCE2 result exhibits a coherence modulation, which is absent in the gCCE3 result. In the main text, we attribute this to multi-spin correlations arising from hyperfine-mediated nuclear spin-spin interactions involving the three nearest neighboring N nuclear spins. Appart from this modulation, we find that the decoherence time converges at the gCCE2 level of theory. At 0.03 T, however, the computed $T_2$ times are 0.15 μs (gCCE2) and 0.10 μs (gCCE3), indicating a non-negligible high-order cluster contribution. We note that higher-order gCCE calculations (e.g. gCCE4) did not yield any converged results due to serious numerical divergence. Thus, it is unclear at which gCCE order the $T_2$ time fully converges. We consider, however, that experimentally measured $T_2$ times in similar magnetic field range from 70 ns to 100 ns[5–8] and prior high-order gCCE calculations reported 92 ns[5], which is close to our gCCE3 result. Thus, we conclude that the $T_2$ time computed at the gCCE2 level of theory is order-of-magnitude correct, with an uncertainty of ~50 ns. We remark, however, that this numerical uncertainty does not affect the main conclusion of our study.

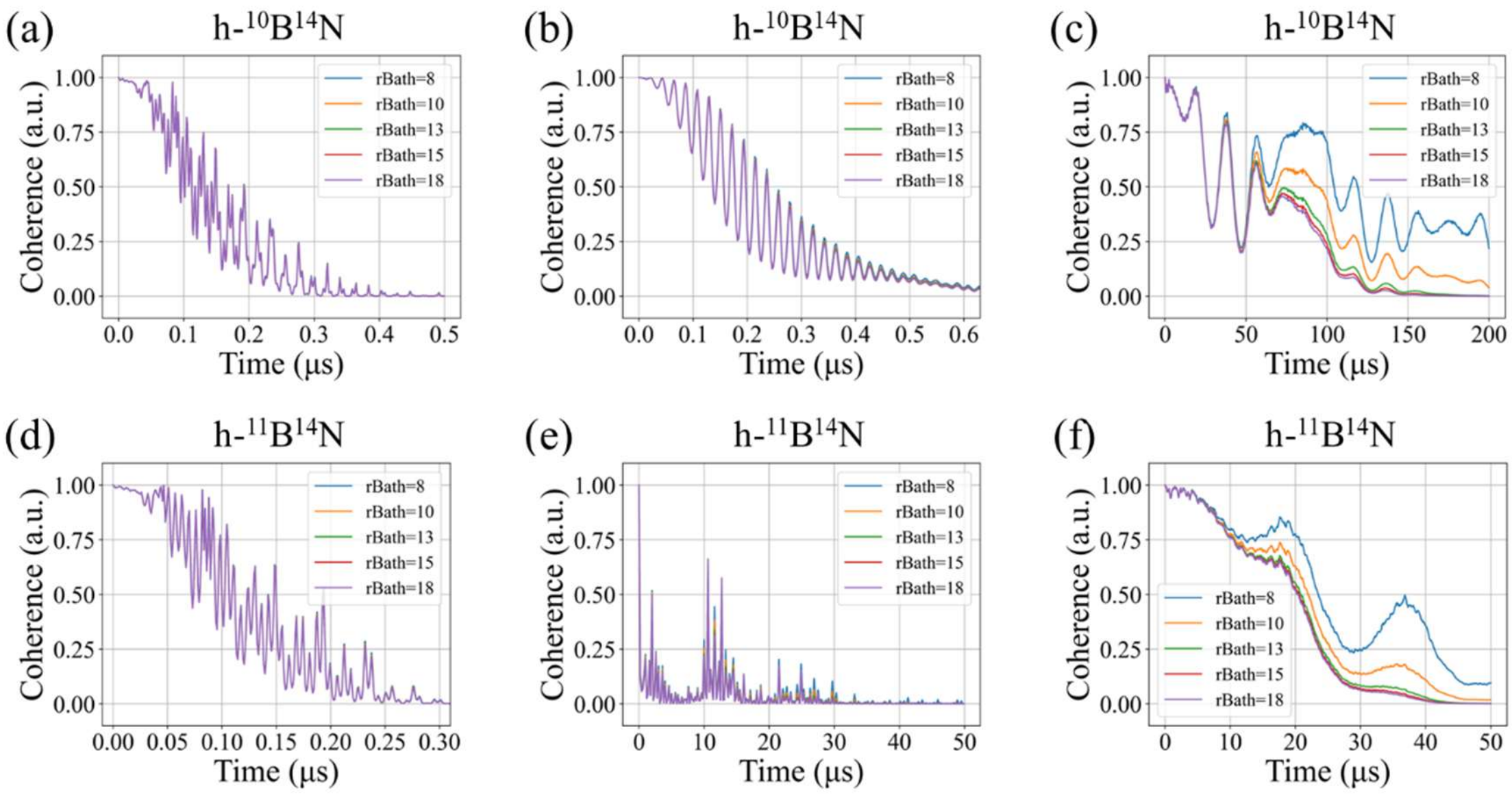

**Figure S1.** Convergence tests of rBath under various magnetic fields and bath configurations by using the gCCE2 method. Magnetic fields are set to 0.03 T (a,d), 0.3 T (b,e), and 3 T (c,f). Bath configurations are h–$^{10}$B$^{14}$N (a,b,c) and h–$^{11}$B$^{14}$N (d,e,f).

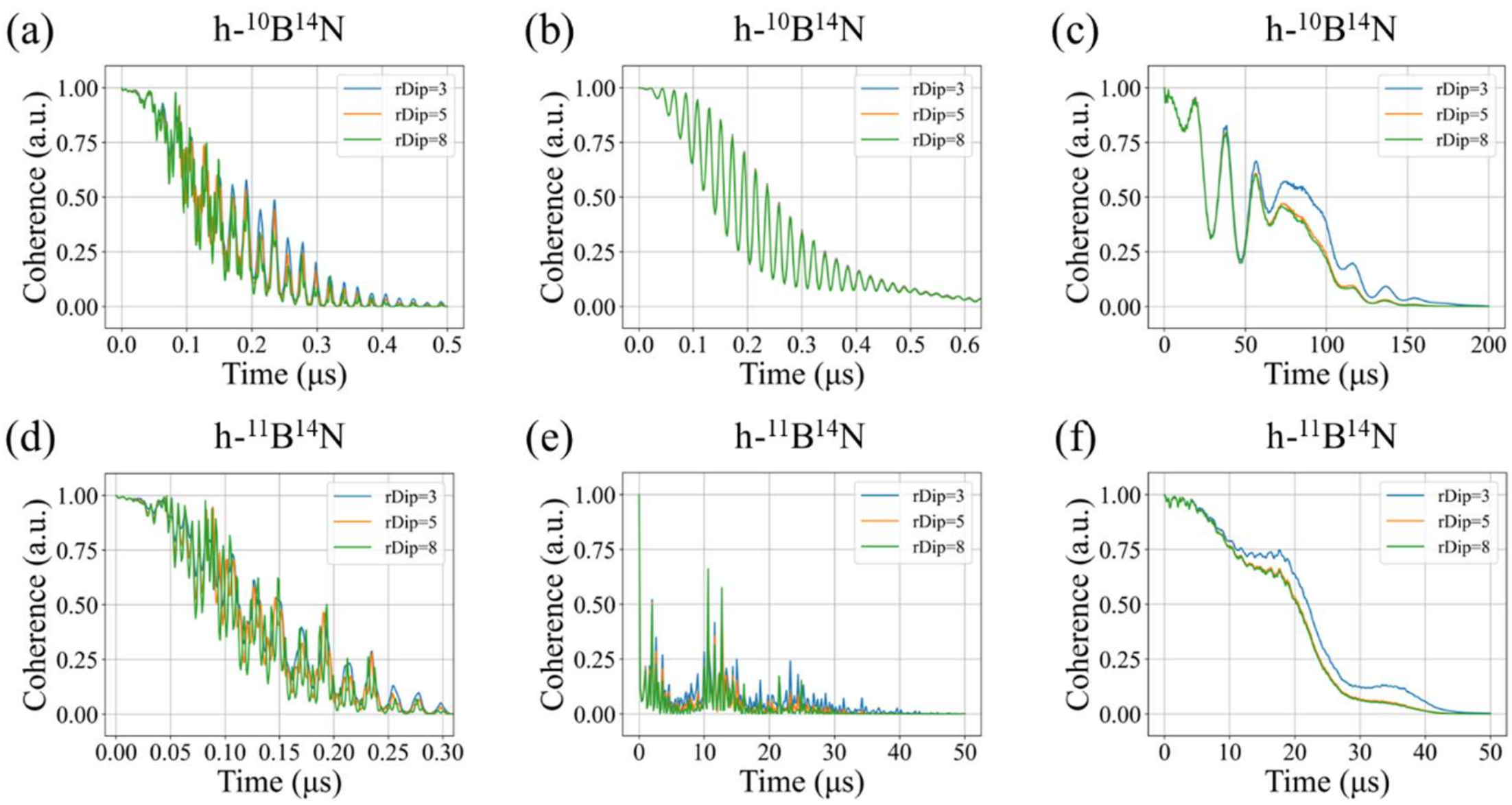

**Figure S2.** Convergence test of rDip under various magnetic fields and bath configurations by using the gCCE2 method. Magnetic fields are set to 0.03 T (a,d), 0.3 T (b,e), and 3 T (c,f). Bath configurations are h–$^{10}$B$^{14}$N (a,b,c) and h–$^{11}$B$^{14}$N (d,e,f).

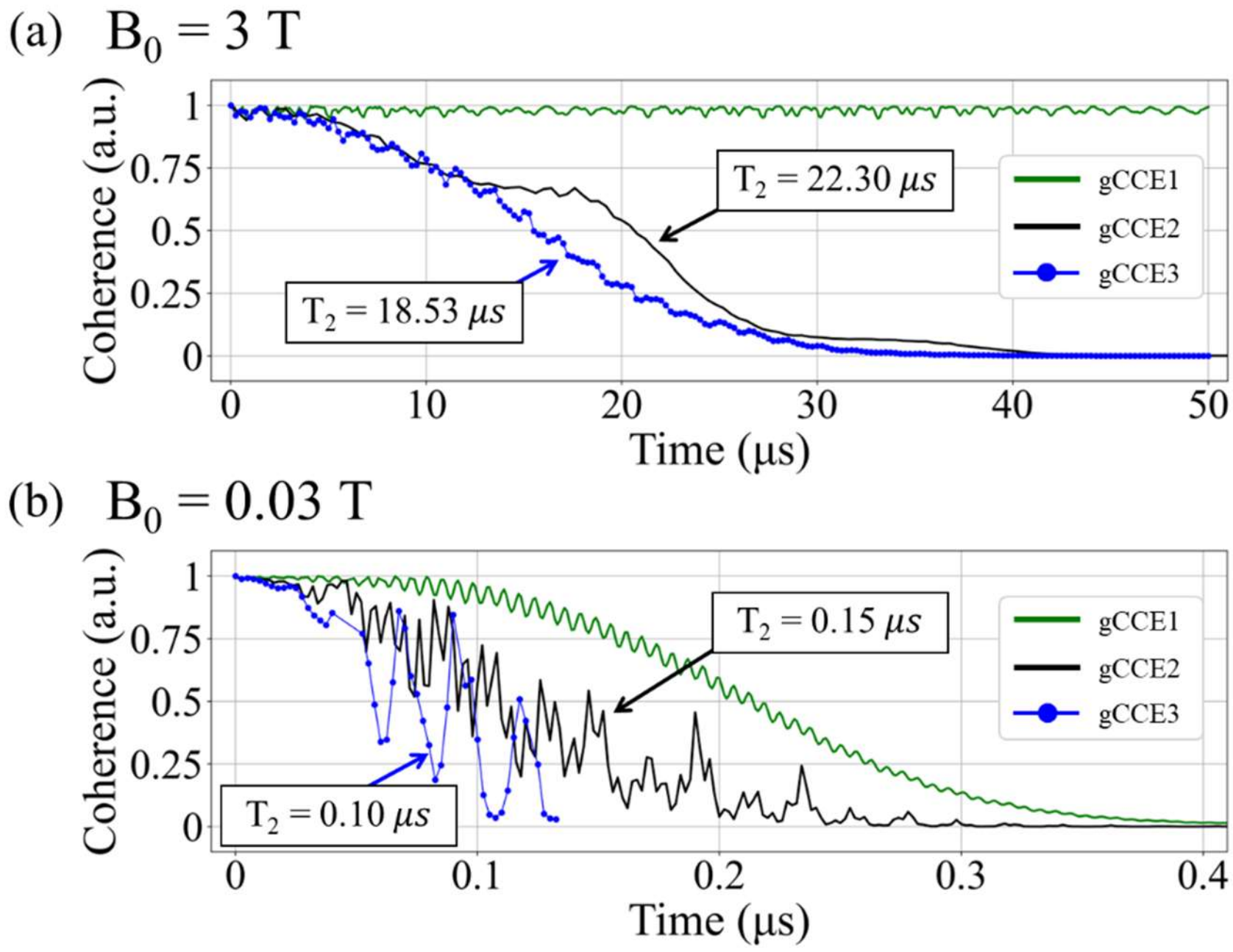

**Figure S3.** $V_B^-$ decoherence computed at different gCCE levels of theory. Coherence under a magnetic field of 3 T (a) and 0.03 T (b). The calculations were conducted for $V_B^-$ in h-$^{11}B^{14}N$, with rBath = 15 Å and rDip = 6 Å.

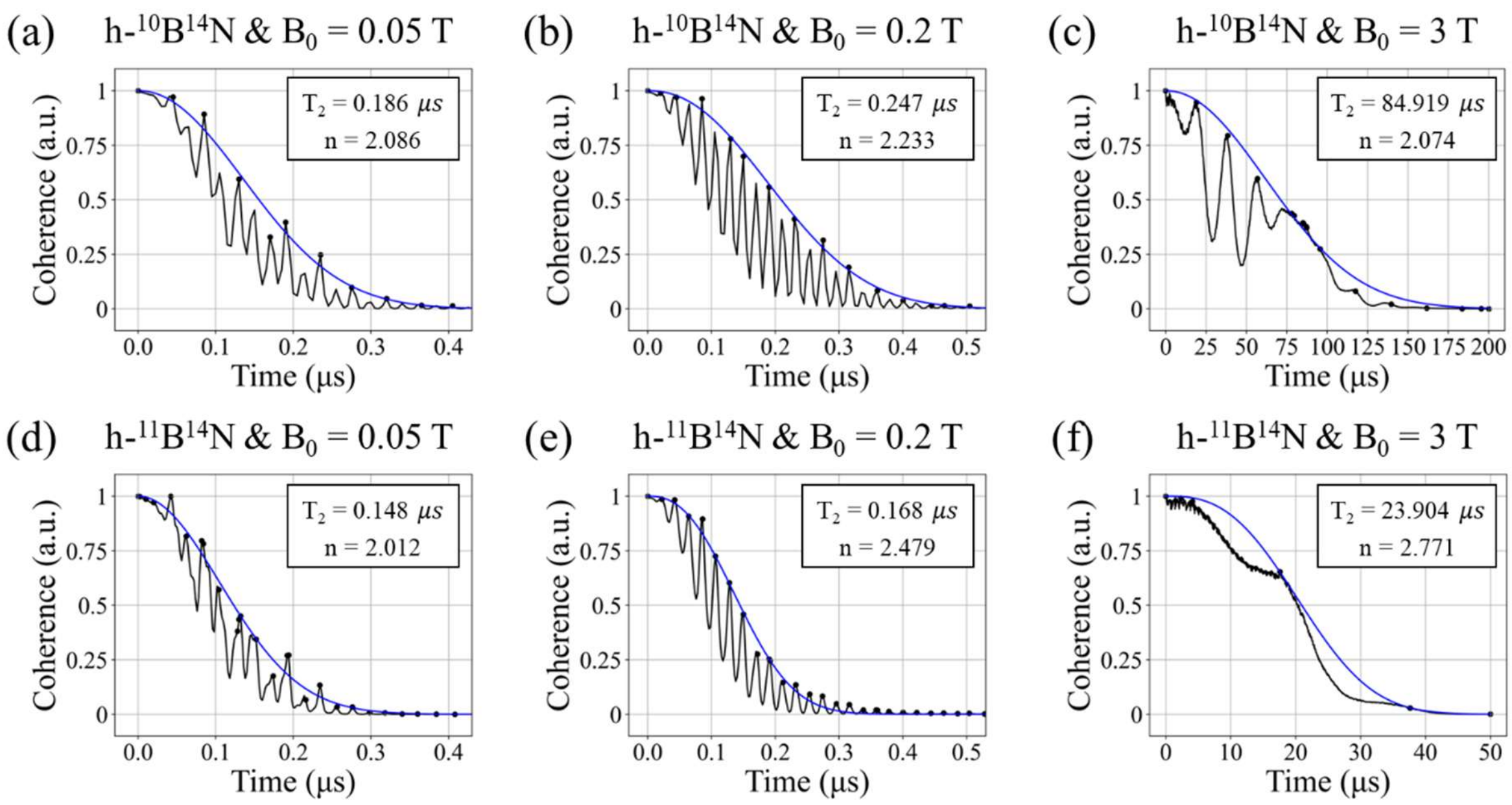

**Figure S4.** Coherence function of the $V_B^-$ spin computed using the gCCE2 method. The coherence function is fitted to a stretched exponential form, $\exp\left[-\left(\frac{2\tau}{T_2}\right)^n\right]$, to compute the $T_2$ time. For the fitting process in the presence of complex modulation feature, we select data points along the upper envelope of the coherence decay and performed the fit. The black dots represent the extracted coherence data for the $T_2$ fitting, and the blue solid lines show the fitted curves using a stretched exponential function. Subfigures (a-c) correspond to h-$^{10}B^{14}N$, and (d-f) to h-$^{11}B^{14}N$. The applied magnetic fields are 0.05 T (a,d), 0.2 T (b,e), and 3 T (c,f), respectively.

**S3. ESEEM equation for an arbitrary nuclear spin**

In this note, we provide the formula for the coherence modulation of the $V_B^-$ spin ($S$ = 1) induced by an arbitrary nuclear spin ($I$). The expression is derived from the coherence function under the Hahn-echo pulse sequence, using the secular approximation as employed in the cCCE method. The total Hamiltonian of the combined $V_B^-$ spin and a nuclear spin system is expressed as follows:[3]

$$H_{tot} = |0\rangle\langle 0| \otimes H^{(|0\rangle)} + |1\rangle\langle 1| \otimes H^{(|1\rangle)} \tag{S11}$$

, where $H^{(|\alpha\rangle)}$ is a nuclear spin Hamiltonian projected on the $|\alpha\rangle$ state of the $V_B^-$ spin. $(|\alpha\rangle = |0\rangle, |1\rangle)$ $H^{(|\alpha\rangle)}$ is given by,

$$H^{(|\alpha\rangle)} = \omega_I \vec{\boldsymbol{I}}_z + \alpha A \vec{\boldsymbol{I}}_z + \alpha B \vec{\boldsymbol{I}}_x + H_Q \tag{S12}$$

, where $\omega_I$ is a nuclear Lamor frequency ($\omega_I = -\gamma_I \hbar B_0$), $A = A_{zz}$ and $B = \sqrt{A_{zx}^2 + A_{zy}^2}$ are secular and pseudo-secular hyperfine interactions, respectively, and $H_Q$ is the quadrupole interaction Hamiltonian. The coherence function of the $V_B^-$ spin in this system is expressed as follows:

$$\mathcal{L}_C(2\tau) = \mathrm{Tr}\left[U_C^{(|0\rangle)}(2\tau)\rho_C U_C^{(|1\rangle)\dagger}(2\tau)\right] \tag{S13}$$

, where $\rho_C$ is the density matrix of the nuclear spin and $U_C^{(|\alpha\rangle)}(2\tau)$ is the time propagator described by the nuclear spin Hamiltonian ($H_C^{(|\alpha\rangle)}$). For the Hahn-echo sequence, the time evolution operator is expressed as,

$$U_C^{(|\alpha\rangle)}(2\tau) = e^{-\frac{i}{\hbar}H_C^{(|\beta\rangle)}\tau} e^{-\frac{i}{\hbar}H_C^{(|\alpha\rangle)}\tau} \tag{S14}$$

, where $|\alpha\rangle$ and $|\beta\rangle$ represent the qubit levels.

To express the formula for the solution, we use eigenvalues and eigenvectors of the nuclear spin Hamiltonian ($H_C^{(|\alpha\rangle)}$), denoted as $\omega_i^{(|\alpha\rangle)}$ and $v_i^{(|\alpha\rangle)}$, respectively, where $i$ is an index satisfying 0 ≤ $I$ ≤ $2I$. The $i$-th nuclear eigenvector of the $|\alpha\rangle$ manifold (e.g. $m_s$ = 1 & $m_I$ = 3/2) is expressed as:

$$v_i^{|\alpha=1\rangle} = c_{i,\frac{3}{2}}^{|\alpha=1\rangle}\left|1,+\frac{3}{2}\right\rangle + c_{i,\frac{1}{2}}^{|\alpha=1\rangle}\left|1,+\frac{1}{2}\right\rangle + c_{i,-\frac{1}{2}}^{|\alpha=1\rangle}\left|1,-\frac{1}{2}\right\rangle + c_{i,-\frac{3}{2}}^{|\alpha=1\rangle}\left|1,-\frac{3}{2}\right\rangle \tag{S15}$$

, where $c_i$ is coefficients of the eigenvector expansion. These eigenvectors and eigenvalues are then substituted into the Equation S14 to derive the coherence function. Consequently, $\mathcal{L}_C(2\tau)$, which is the ESEEM formula for a single nuclear spin number ($I$) is expressed as follows[9–11]:

$$\mathcal{L}_C(2\tau) = \frac{1}{2I+1}\left(\sum_{i,k=0}^{2I}|M_{ik}|^4 + 2\sum_{j=1}^{2I}\sum_{i=0}^{j-1}\sum_{k=0}^{2I}\cos\left(\omega_{ij}^{(|\alpha\rangle)}\tau\right)|M_{ik}|^2|M_{jk}|^2 \right.$$
$$+ 2\sum_{n=1}^{2I}\sum_{k=0}^{n-1}\sum_{i=0}^{2I}\cos\left(\omega_{kn}^{(|\beta\rangle)}\tau\right)|M_{ik}|^2|M_{in}|^2$$
$$+ 2\sum_{j=1}^{2I}\sum_{i=0}^{j-1}\sum_{n=1}^{2I}\sum_{k=0}^{n-1}\cos\left(\left(\omega_{ij}^{(|\alpha\rangle)} - \omega_{kn}^{(|\beta\rangle)}\right)\tau\right)M_{in}^*M_{jk}^*M_{ik}M_{jn}$$
$$\left. + 2\sum_{j=1}^{2I}\sum_{i=0}^{j-1}\sum_{n=1}^{2I}\sum_{k=0}^{n-1}\cos\left(\left(\omega_{ij}^{(|\alpha\rangle)} + \omega_{kn}^{(|\beta\rangle)}\right)\tau\right)M_{ik}^*M_{jn}^*M_{in}M_{jk}\right) \quad \text{(S16)}$$

, where $i$ and $j$ refer to the indices of the $i$-th and $j$-th sublevels within the $|\alpha\rangle$ state manifold, and $n$ and $k$ refer to the indices the $k$-th and $n$-th sublevels within the $|\beta\rangle$ state manifold. The nuclear transition frequency ($\omega_{ij}^{(|\alpha\rangle)}$) is derived from the eigenvalues:

$$\omega_{ij}^{(|\alpha\rangle)} = \omega_i^{(|\alpha\rangle)} - \omega_j^{(|\alpha\rangle)} \quad \text{(S17)}$$

Similarly, $\omega_{ij}^{(|\beta\rangle)}$ is derived from the eigenvalues within to the $|\beta\rangle$ manifold. The matrix element ($M_{ik}$) is derived from the eigenvalues and eigenvectors of the nuclear spin Hamiltonian:

$$M_{ik} = \vec{\boldsymbol{v}}_i^{*(|\alpha\rangle)} \cdot \vec{\boldsymbol{v}}_k^{(|\beta\rangle)} = \sum_l v_{i,l}^{*(|\alpha\rangle)} v_{k,l}^{(|\beta\rangle)} \quad \text{(S18)}$$

, where $\vec{\boldsymbol{v}}_i^*$ is the complex conjugate of the eigenvector for the $i$-th sublevel, and $v_{i,l}$ is the $l$-th element of the eigenvector for the $i$-th sublevel. The matrix gives the electron paramagnetic resonance (EPR) transition amplitude from the $i$-th sublevel within the $|\alpha\rangle$ state manifold to the $k$-th sublevel within the $|\beta\rangle$ state manifold. This ESEEM equation can be divided into five terms[11],

$$\mathcal{L}_C(2\tau) = V_0 + V_\alpha(2\tau) + V_\beta(2\tau) + V_-(2\tau) + V_+(2\tau) \quad \text{(S19)}$$

In Equation S19, the un-modulation part ($V_0$) is independent of time and is composed solely of matrix elements $M_{ik}$:

$$V_0 = \frac{1}{2I+1}\sum_{i=0}^{2I}\sum_{k=0}^{2I}|M_{ik}|^4 \quad \text{(S20)}$$

The modulations with nuclear frequencies in $|\alpha\rangle$ and $|\beta\rangle$ state manifold are:

$$V_\alpha(2\tau) = \frac{2}{2I+1}\sum_{j=1}^{2I}\sum_{i=0}^{j-1}\sum_{k=0}^{2I}\cos\left(\omega_{ij}^{(|\alpha\rangle)}\tau\right)|M_{ik}|^2|M_{jk}|^2 \tag{S21}$$

$$V_\beta(2\tau) = \frac{2}{2I+1}\sum_{n=1}^{2I}\sum_{k=0}^{n-1}\sum_{i=0}^{2I}\cos\left(\omega_{kn}^{(|\beta\rangle)}\tau\right)|M_{ik}|^2|M_{in}|^2 \tag{S22}$$

The modulation with difference combination frequencies is:

$$V_-(2\tau) = \frac{2}{2I+1}\sum_{j=1}^{2I}\sum_{i=0}^{j-1}\sum_{n=1}^{2I}\sum_{k=0}^{n-1}\cos\left(\left(\omega_{ij}^{(|\alpha\rangle)} - \omega_{kn}^{(|\beta\rangle)}\right)\tau\right)M_{in}^*M_{jk}^*M_{ik}M_{jn} \tag{S23}$$

The modulation with sum combination frequencies is:

$$V_+(2\tau) = \frac{2}{2I+1}\sum_{j=1}^{2I}\sum_{i=0}^{j-1}\sum_{n=1}^{2I}\sum_{k=0}^{n-1}\cos\left(\left(\omega_{ij}^{(|\alpha\rangle)} + \omega_{kn}^{(|\beta\rangle)}\right)\tau\right)M_{ik}^*M_{jn}^*M_{in}M_{jk} \tag{S24}$$

Figure S5 compares the results of the cCCE1 method and the modulation parts of ESEEM equation for a combined $V_B^-$ spin and a nearest neighboring $^{14}N$ spin system. We find that these methods yield the same result, confirming the validity of the ESEEM formula. In addition, we note that the ESEEM formula provides a way to decompose the coherence into the unmodulated and modulated terms, enabling detailed analysis on the nuclear modulation effect. Figure S5c shows the FFT spectrum of the coherence, exhibiting three frequency components. The ESEEM formula also explains the origin of these frequencies.

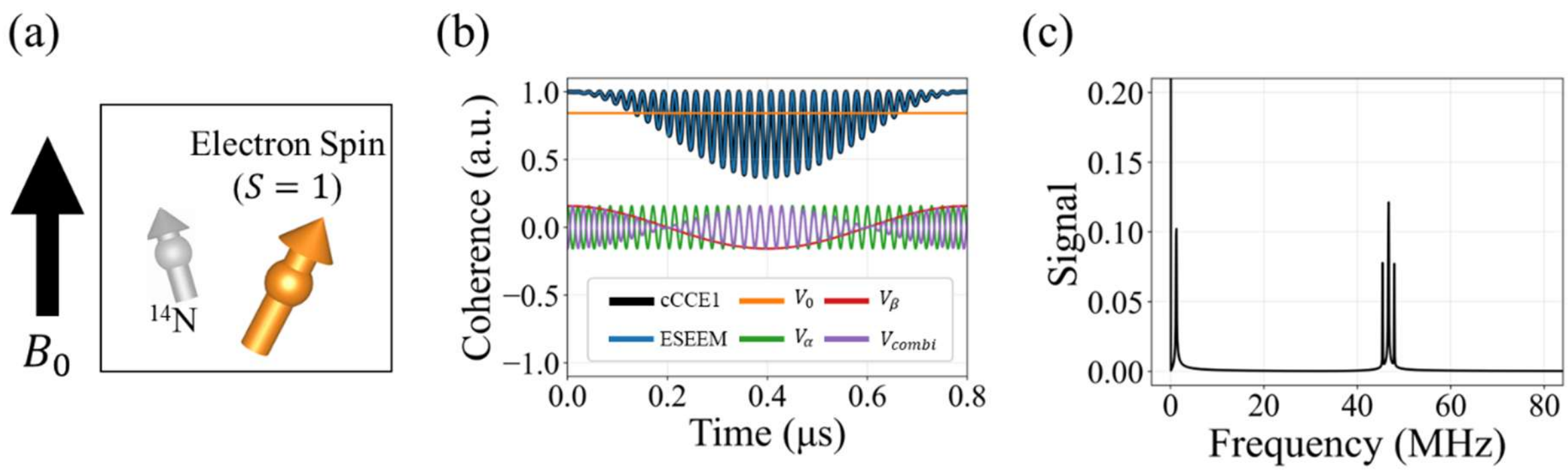

**Figure S5. a)** Schematic of a nearest neighboring 14N nuclear spin and a $V_B^-$ electron spin. **b)** Computed coherence of the $V_B^-$ spin using the ESEEM formula and the cCCE method at $B_0$ = 0.3 T; the black line represents the cCCE method, the blue line represents the ESEEM equation, the orange line corresponds to $V_0$, the green line to $V_\alpha$, the red line to $V_\beta$, purple line to $V_{combi}(= V_+ + V_-)$. **c)** FFT spectrum of the coherence at $B_0$ = 0.3 T.

## S4. High-frequency coherence modulation of $V_B^-$ due to the $^{14}N$ nuclear spin

To identify the high-frequency modulation associated with the nearest neighboring $^{14}N$ nuclear spin, we analytically solve the ESEEM equation. The nuclear spin Hamiltonian for the nearest $^{14}N$, described in Equation S12, can be represented in matrix form as follows:

$$H^{(|\alpha\rangle)} = \omega_I \begin{pmatrix} 1 & 0 & 0 \\ 0 & 0 & 0 \\ 0 & 0 & -1 \end{pmatrix} + \alpha A_{zz} \begin{pmatrix} 1 & 0 & 0 \\ 0 & 0 & 0 \\ 0 & 0 & -1 \end{pmatrix} + \frac{\alpha A_{zx}}{\sqrt{2}} \begin{pmatrix} 0 & 1 & 0 \\ 1 & 0 & 1 \\ 0 & 1 & 0 \end{pmatrix} + \frac{A_{zy}}{\sqrt{2}} \begin{pmatrix} 0 & -i & 0 \\ i & 0 & -i \\ 0 & i & 0 \end{pmatrix} + H_Q \quad \text{(S25)}$$

, where $i$ represents imaginary number and $A_{ab}$ represents hyperfine tensor parameters. In Equation S25, the pseudo-secular hyperfine parameter $A_{zx}$ and $A_{zy}$ are zero because the nearest neighboring $^{14}N$ nuclear spin is in the plane. The quadrupole tensor $\overleftrightarrow{\mathbf{Q}}$ is computed from the second partial derivative of the electric potential $V$, given by[4]:

$$Q_{ab} = \frac{eQ}{2I(2I+1)\hbar} \frac{\partial^2 V}{\partial x_a \partial x_b} \quad \text{(S26)}$$

, where $eQ$ is the nuclear quadrupole moment and $\hbar$ is the reduced Planck constant.

Due to the symmetry of the electric potential in the plane of the $V_B^-$ defect (xy-plane), the quadrupole tensor elements $Q_{zx}$ and $Q_{zy}$ vanish for nuclear spins located in the plane. The quadrupole Hamiltonian of the nearest neighboring $^{14}N$ nuclear spin is written as follows:

$$\begin{aligned} H_Q &= Q_{xx}\vec{\boldsymbol{I}}_x\vec{\boldsymbol{I}}_x + Q_{yy}\vec{\boldsymbol{I}}_y\vec{\boldsymbol{I}}_y + Q_{zz}\vec{\boldsymbol{I}}_z\vec{\boldsymbol{I}}_z + Q_{xy}\vec{\boldsymbol{I}}_x\vec{\boldsymbol{I}}_y + Q_{yx}\vec{\boldsymbol{I}}_y\vec{\boldsymbol{I}}_x \\ &= \frac{Q_{xx}}{2}\begin{pmatrix} 1 & 0 & 1 \\ 0 & 2 & 0 \\ 1 & 0 & 1 \end{pmatrix} + \frac{Q_{yy}}{2}\begin{pmatrix} 1 & 0 & -1 \\ 0 & 2 & 0 \\ -1 & 0 & 1 \end{pmatrix} + Q_{zz}\begin{pmatrix} 1 & 0 & 0 \\ 0 & 0 & 0 \\ 0 & 0 & 1 \end{pmatrix} \\ &\quad + \frac{Q_{xy}}{2}\begin{pmatrix} i & 0 & -i \\ 0 & 0 & 0 \\ i & 0 & -i \end{pmatrix} + \frac{Q_{yx}}{2}\begin{pmatrix} -i & 0 & -i \\ 0 & 0 & 0 \\ i & 0 & i \end{pmatrix} \end{aligned} \quad \text{(S27)}$$

, where $Q_{ab}$ represents the quadrupole tensor parameters. In Equation S27, the $Q_{xy}$ and $Q_{yx}$ are the same. The spin Hamiltonian can be rewritten in the following form:

$$\begin{aligned} H^{(|\alpha\rangle)} &= \begin{pmatrix} \omega_I + \alpha A_{zz} + \frac{Q_{xx}+Q_{yy}}{2} + Q_{zz} & 0 & \frac{Q_{xx}-Q_{yy}}{2} - iQ_{xy} \\ 0 & Q_{xx}+Q_{yy} & 0 \\ \frac{Q_{xx}-Q_{yy}}{2} + iQ_{xy} & 0 & -\omega_I - \alpha A_{zz} + \frac{Q_{xx}+Q_{yy}}{2} + Q_{zz} \end{pmatrix} \\ &= \begin{pmatrix} a_{11}^{(|\alpha\rangle)} & 0 & a_{13} - ib_{13} \\ 0 & a_{22}^{(|\alpha\rangle)} & 0 \\ a_{13} + ib_{13} & 0 & a_{33}^{(|\alpha\rangle)} \end{pmatrix} \end{aligned} \quad \text{(S28)}$$

The eigenvalues of the effective Hamiltonian in Equation S28 are

$$\lambda_1^{(|\alpha\rangle)} = \frac{a_{11}^{(|\alpha\rangle)} + a_{33}^{(|\alpha\rangle)}}{2} + \frac{1}{2}\sqrt{\left(a_{11}^{(|\alpha\rangle)} - a_{33}^{(|\alpha\rangle)}\right)^2 + 4a_{13} + 4b_{13}}$$

$$= \frac{Q_{xx} + Q_{yy}}{2} + Q_{zz} + \sqrt{(\omega_I + \alpha A_{zz})^2 + \left(\frac{Q_{xx} - Q_{yy}}{2}\right)^2 + Q_{xy}^2} \qquad \text{(S29)}$$

$$\lambda_2^{(|\alpha\rangle)} = a_{22}^{(|\alpha\rangle)} = Q_{xx} + Q_{yy} \qquad \text{(S30)}$$

$$\lambda_3^{(|\alpha\rangle)} = \frac{a_{11}^{(|\alpha\rangle)} + a_{33}^{(|\alpha\rangle)}}{2} - \frac{1}{2}\sqrt{\left(a_{11}^{(|\alpha\rangle)} - a_{33}^{(|\alpha\rangle)}\right)^2 + 4a_{13} + 4b_{13}}$$

$$= \frac{Q_{xx} + Q_{yy}}{2} + Q_{zz} - \sqrt{(\omega_I + \alpha A_{zz})^2 + \left(\frac{Q_{xx} - Q_{yy}}{2}\right)^2 + Q_{xy}^2} \qquad \text{(S31)}$$

The eigenvectors of the effective Hamiltonian in Equation S28 are

$$\nu_1^{(|\alpha\rangle)} = \left\{\frac{a_{11}^{(|\alpha\rangle)} - a_{33}^{(|\alpha\rangle)} + \sqrt{\left(a_{11}^{(|\alpha\rangle)} - a_{33}^{(|\alpha\rangle)}\right)^2 + 4a_{13} + 4b_{13}}}{2(a_{13} + ib_{13})}, 0, 1\right\} \qquad \text{(S32)}$$

$$\nu_2^{(|\alpha\rangle)} = \{0,1,0\} \qquad \text{(S33)}$$

$$\nu_3^{(|\alpha\rangle)} = \left\{\frac{a_{11}^{(|\alpha\rangle)} - a_{33}^{(|\alpha\rangle)} - \sqrt{\left(a_{11}^{(|\alpha\rangle)} - a_{33}^{(|\alpha\rangle)}\right)^2 + 4a_{13} + 4b_{13}}}{2(a_{13} + ib_{13})}, 0, 1\right\} \qquad \text{(S34)}$$

The nuclear transition frequencies in the $V_B^-$ spin's $|\alpha\rangle$ manifold are

$$\omega_{12}^{(|\alpha\rangle)} = \lambda_1^{(|\alpha\rangle)} - \lambda_2^{(|\alpha\rangle)} = -\frac{Q_{xx} + Q_{yy}}{2} + Q_{zz} + \sqrt{(\omega_I + \alpha A_{zz})^2 + \left(\frac{Q_{xx} - Q_{yy}}{2}\right)^2 + Q_{xy}^2} \qquad \text{(S35)}$$

$$\omega_{23}^{(|\alpha\rangle)} = \lambda_2^{(|\alpha\rangle)} - \lambda_3^{(|\alpha\rangle)} = \frac{Q_{xx} + Q_{yy}}{2} - Q_{zz} + \sqrt{(\omega_I + \alpha A_{zz})^2 + \left(\frac{Q_{xx} - Q_{yy}}{2}\right)^2 + Q_{xy}^2} \qquad \text{(S36)}$$

$$\omega_{13}^{(|\alpha\rangle)} = \lambda_1^{(|\alpha\rangle)} - \lambda_3^{(|\alpha\rangle)} = 2\sqrt{(\omega_I + \alpha A_{zz})^2 + \left(\frac{Q_{xx} - Q_{yy}}{2}\right)^2 + Q_{xy}^2} \qquad \text{(S37)}$$

These nuclear transition frequencies contribute to the modulation caused by the nearest neighboring $^{14}$N nuclear spins, as described in Equation S21-S24, yielding theoretically up to 24 possible peak frequencies in the Fourier transform. However, as shown in Figure S5c, only four frequencies are observed in the modulation caused by the nearest neighboring $^{14}$N nuclear spin. This is because $M_{2i} = M_{i2} = 0$ ($i$ = 1,3) eliminates the cosine terms associated with $\omega_{12}$ and $\omega_{23}$ in Equation S21-S24. The four peak frequencies identified in the FFT spectrum are 1.3, 45.4, 46.7, and 47.9 MHz, corresponding to $\omega_{13}^{(|0\rangle)}/2$ , $(\omega_{13}^{(|1\rangle)} - \omega_{13}^{(|0\rangle)})/2$, $\omega_{13}^{(|1\rangle)}/2$, and $(\omega_{13}^{(|1\rangle)} + \omega_{13}^{(|0\rangle)})/2$, respectively. We identify that the high-frequency modulation caused by the nearest $^{14}N$, which are shown in Figure 2c of the main text, arises from the modulation frequencies $\omega_{13}^{(|0\rangle)}/2$ and $\omega_{13}^{(|1\rangle)}/2$.

**S5. The cancellation condition for arbitrary nuclear spin**

The "cancellation condition" referenced in our work originates from earlier EPR studies, where it is commonly referred to as the "exact cancellation condition".[9,10,12] In those works, this condition arises when the nuclear Zeeman energy and the hyperfine interaction cancel each other, leading to a significant increase in ESEEM depth and the appearance of sharp, narrow peaks in its Fourier spectrum.

For a system with an electron spin $S$ = 1/2 and nuclear spin $I$ = 1/2, the exact cancellation condition is typically defined as:

$$2\omega_I = A$$

, where $\omega_I$ is the nuclear Larmore frequency and $A$ is the longitudinal hyperfine component. Under this condition, level anti-crossing occurs between nuclear spin sublevels, which maximizes the modulation depth. The depth parameter $k$, which quantifies this modulation, has an analytical form:

$$k = \frac{B^2\omega_I^2}{\left(\left(\omega_I + \frac{A}{2}\right)^2 + \left(\frac{B}{2}\right)^2\right)\left(\left(\omega_I - \frac{A}{2}\right)^2 + \left(-\frac{B}{2}\right)^2\right)}$$

, which $B$ represents the transverse hyperfine component. The parameter $k$ reaches its maximum when $\omega_I = A/2$, corresponding to the exact cancellation condition. In addition to this exact cancellation condition, a near cancellation condition can also be defined.[13] This refers to the case in which the nuclear Zeeman energy is nearly – but not exactly – canceled by the hyperfine interaction. While the modulation depth in this case does not reach the theoretical maximum, pronounced ESEEM features and fluctuations in modulation depth may still be observed.

In our work, we extend the concept of the cancellation condition beyond spin-1/2 systems to arbitrary nuclear spins (e.g., $I$ = 3 or 3/2). Through generalized analysis of the ESEEM expression, we show that strong modulation arises whenever level anti-crossings occur between nuclear spin sublevels. To capture this behavior in high spin systems, we analyze the ratio $|\Omega/\Delta|$, where $\Omega$ is the coupling rate between nuclear spin sublevels and $\Delta$ is their energy splitting. Sharp peaks in this ratio indicate level anti-crossings and correspond to strong nuclear modulation. To illustrate the relationship between level anti-crossing and strong modulation, we present Figure S6, which shows the energy level diagrams, $|\Omega/\Delta|$ ratio, and ESEEM modulation as function of magnetic field for nuclear spin with $I = 3/2$. In particular, for a $V_B$- defect coupled to $^{11}B$ with $I = 3/2$, we observe level anti-crossings at approximately 0.05, 0.12, and 0.18 T. These points coincide with sharp peaks in the $|\Omega/\Delta|$ ratio and strong modulation. This relationship confirms that the cancellation condition provides a microscopic explanation for the strong decoherence observed below the TB in Figure 4 of the main manuscript.

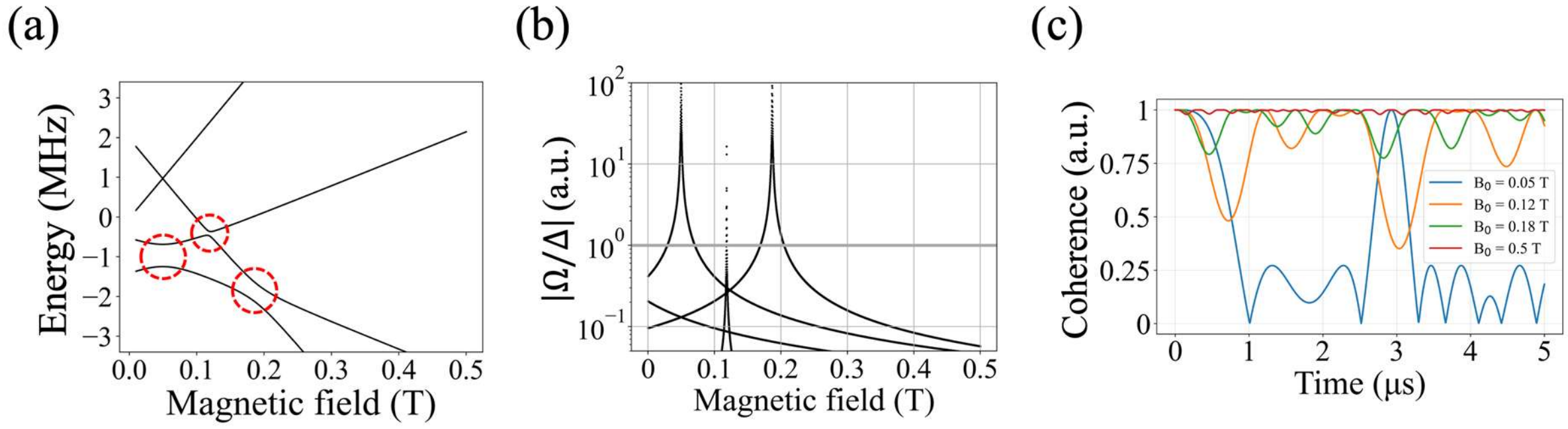

**Figure S6. a)** Energy level diagram of a single $^{11}B$ nuclear spin in h-BN under the hyperfine field of the $V_B^-$ spin in the $m_s$=+1 state. Red circles indicate level anti-crossing points, which occur at magnetic field strengths of 0.05, 0.12, and 0.18 T. **b)** The ratio $|\Omega/\Delta|$ for the $^{11}B$ nuclear spin as a function of magnetic field strength. **c)** ESEEM modulation of the $V_B^-$ spin due to the $^{11}B$ nuclear spin at magnetic fields corresponding to the level anti-crossing points (0.05, 0.12, and 0.18 T), and at 0.5 T, which is away from the level anti-crossing condition.

**S6. The cancellation condition for boron nuclear spins in $V_B^-$ defect**

For cases where the nuclear spin is larger than 1, such as boron isotopes, the cancellation condition has not been well investigated. In this study, we examine the cancellation condition for boron isotopes by considering the ratio of the coupling rate (Ω) to energy level difference (Δ) between two nuclear spin states, which determines the conditions under which the modulation depth can increase. The energy levels and the coupling rates are derived from the nuclear spin Hamiltonian projected on an electron spin state given by Equation S12. When the ratio ($|\Omega/\Delta|$) exceeds 1, the modulation depths in Equation S21-S24 become significantly larger. We also note that for an arbitrary nuclear spin ($I$), the number of possible cancellation conditions is $4I$ - 1, which is the sum of the number of possible single-quantum transitions ($2I$) and double-quantum transitions ($2I$ - 1). For instance, in the case of $^{11}B$ ($I$ = 3/2), there are three possible single-quantum transitions, and their cancellation conditions are determined by the following ratios:

$$\frac{\Omega_{\left(\frac{3}{2},\frac{1}{2}\right)}}{\Delta_{\left(\frac{3}{2},\frac{1}{2}\right)}}=\frac{\frac{\sqrt{3}}{2}m_s\left(A_{zx}-i\cdot A_{zy}\right)+\sqrt{3}\left(Q_{xz}-i\cdot Q_{yz}\right)}{\omega_I+m_sA_{zz}-Q_{xx}-Q_{yy}+2Q_{zz}} \quad \text{(S38)}$$

$$\frac{\Omega_{\left(\frac{1}{2},-\frac{1}{2}\right)}}{\Delta_{\left(\frac{1}{2},-\frac{1}{2}\right)}}=\frac{m_s\left(A_{zx}-i\cdot A_{zy}\right)}{\omega_I+m_sA_{zz}} \quad \text{(S39)}$$

$$\frac{\Omega_{\left(-\frac{1}{2},-\frac{3}{2}\right)}}{\Delta_{\left(-\frac{1}{2},-\frac{3}{2}\right)}}=\frac{\frac{\sqrt{3}}{2}m_s\left(A_{zx}-i\cdot A_{zy}\right)-\sqrt{3}\left(Q_{xz}-i\cdot Q_{yz}\right)}{\omega_I+m_sA_{zz}+Q_{xx}+Q_{yy}-2Q_{zz}} \quad \text{(S40)}$$

, where $\Omega_{(i,j)}$ and $\Delta_{(i,j)}$ represent the coupling rate and the energy difference between two different states, $i$ and $j$. The $m_s$ denotes the electron spin projection (which can be $\alpha$ or $\beta$) and $\omega_I$ is a Lamor frequency of nuclear spin ($\omega_I=-\gamma_I\hbar B_0$). In addition, there are two possible double-quantum transitions due to the quadrupole interaction, and their cancellation conditions are determined by the following ratios:

$$\frac{\Omega_{\left(\frac{3}{2},-\frac{1}{2}\right)}}{\Delta_{\left(\frac{3}{2},-\frac{1}{2}\right)}}=\frac{\frac{\sqrt{3}}{2}\left(Q_{xx}-Q_{yy}\right)-\sqrt{3}i\cdot Q_{\mathrm{xy}}}{2\omega_I+2m_sA_{zz}-Q_{xx}-Q_{yy}+2Q_{zz}} \quad \text{(S41)}$$

$$\frac{\Omega_{\left(-\frac{1}{2},-\frac{3}{2}\right)}}{\Delta_{\left(-\frac{1}{2},-\frac{3}{2}\right)}}=\frac{\frac{\sqrt{3}}{2}\left(Q_{xx}-Q_{yy}\right)-\sqrt{3}i\cdot Q_{\mathrm{xy}}}{2\omega_I+2m_sA_{zz}+Q_{xx}+Q_{yy}-2Q_{zz}} \quad \text{(S42)}$$

In the case of $^{10}$B ($I$ = 3), there are eleven possible channels for the cancellation condition, consisting of six single-quantum transitions and five double-quantum transitions. The six cancellation conditions for single-quantum transition are determined by the following ratios:

$$\frac{\Omega_{(3,2)}}{\Delta_{(3,2)}} = \frac{\frac{\sqrt{6}}{2} m_s\left(A_{zx} - i \cdot A_{zy}\right) + \frac{5\sqrt{6}}{2}\left(Q_{xz} - i \cdot Q_{yz}\right)}{\omega_I + m_s A_{zz} - \frac{5}{2}\left(Q_{xx} + Q_{yy}\right) + 5Q_{zz}} \tag{S43}$$

$$\frac{\Omega_{(2,1)}}{\Delta_{(2,1)}} = \frac{\frac{\sqrt{10}}{2} m_s\left(A_{zx} - i \cdot A_{zy}\right) + \frac{3\sqrt{10}}{2}\left(Q_{xz} - i \cdot Q_{yz}\right)}{\omega_I + m_s A_{zz} - \frac{3}{2}\left(Q_{xx} + Q_{yy}\right) + 3Q_{zz}} \tag{S44}$$

$$\frac{\Omega_{(1,0)}}{\Delta_{(1,0)}} = \frac{\sqrt{3} m_s\left(A_{zx} - i \cdot A_{zy}\right) + \sqrt{3}\left(Q_{xz} - i \cdot Q_{yz}\right)}{\omega_I + m_s A_{zz} - \frac{1}{2}\left(Q_{xx} + Q_{yy}\right) + Q_{zz}} \tag{S45}$$

$$\frac{\Omega_{(0,-1)}}{\Delta_{(0,-1)}} = \frac{\sqrt{3} m_s\left(A_{zx} - i \cdot A_{zy}\right) - \sqrt{3}\left(Q_{xz} - i \cdot Q_{yz}\right)}{\omega_I + m_s A_{zz} + \frac{1}{2}\left(Q_{xx} + Q_{yy}\right) - Q_{zz}} \tag{S46}$$

$$\frac{\Omega_{(-1,-2)}}{\Delta_{(-1,-2)}} = \frac{\frac{\sqrt{10}}{2} m_s\left(A_{zx} - i \cdot A_{zy}\right) - \frac{3\sqrt{10}}{2}\left(Q_{xz} - i \cdot Q_{yz}\right)}{\omega_I + m_s A_{zz} + \frac{3}{2}\left(Q_{xx} + Q_{yy}\right) - 3Q_{zz}} \tag{S47}$$

$$\frac{\Omega_{(-2,-3)}}{\Delta_{(-2,-3)}} = \frac{\frac{\sqrt{6}}{2} m_s\left(A_{zx} - i \cdot A_{zy}\right) - \frac{5\sqrt{6}}{2}\left(Q_{xz} - i \cdot Q_{yz}\right)}{\omega_I + m_s A_{zz} + \frac{5}{2}\left(Q_{xx} + Q_{yy}\right) - 5Q_{zz}} \tag{S48}$$

Similarly, the cancellation conditions for double-quantum transitions are determined by the following ratios:

$$\frac{\Omega_{(3,1)}}{\Delta_{(3,1)}} = \frac{\frac{\sqrt{15}}{2}\left(Q_{xx} - Q_{yy}\right) - i \cdot \sqrt{15} Q_{xy}}{2\omega_I + 2m_s A_{zz} - 4\left(Q_{xx} + Q_{yy}\right) + 8Q_{zz}} \tag{S49}$$

$$\frac{\Omega_{(2,0)}}{\Delta_{(2,0)}} = \frac{\frac{\sqrt{30}}{2}\left(Q_{xx} - Q_{yy}\right) - i \cdot \sqrt{30} Q_{xy}}{2\omega_I + 2m_s A_{zz} - 2\left(Q_{xx} + Q_{yy}\right) + 4Q_{zz}} \tag{S50}$$

$$\frac{\Omega_{(1,-1)}}{\Delta_{(1,-1)}} = \frac{3\left(Q_{xx} - Q_{yy}\right) - i \cdot 6 Q_{xy}}{2\omega_I + 2m_s A_{zz}} \tag{S51}$$

$$\frac{\Omega_{(0,-2)}}{\Delta_{(0,-2)}} = \frac{\frac{\sqrt{30}}{2}\left(Q_{xx} - Q_{yy}\right) - i \cdot \sqrt{30} Q_{xy}}{2\omega_I + 2m_s A_{zz} + 2\left(Q_{xx} + Q_{yy}\right) - 4Q_{zz}} \tag{S52}$$

$$\frac{\Omega_{(-1,-3)}}{\Delta_{(-1,-3)}} = \frac{\frac{\sqrt{15}}{2}\left(Q_{xx} - Q_{yy}\right) - i \cdot \sqrt{15} Q_{xy}}{2\omega_I + 2m_s A_{zz} + 4\left(Q_{xx} + Q_{yy}\right) - 8Q_{zz}} \tag{S53}$$

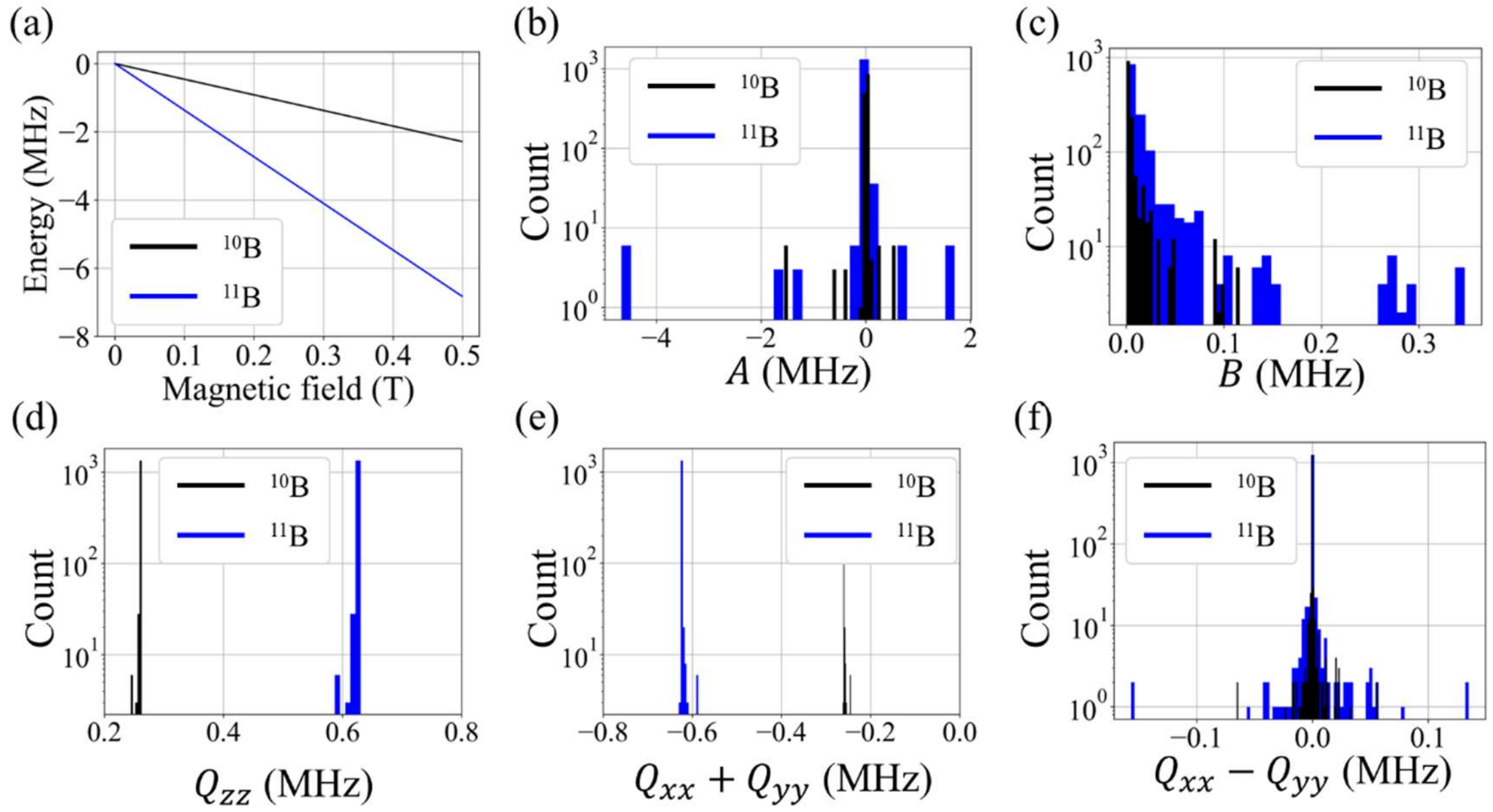

**Figure S7. a)** Zeeman energy of boron nuclear spins as a function of magnetic field. **b,c)** Histogram of the hyperfine coupling constants $A\ (= A_{zz})$ and $B\ (= \sqrt{A_{zx}^2 + A_{zy}^2})$ for boron nuclear spins in h-BN with $V_B^-$ defect: secular (b) and pseudo-secular (c) hyperfine couplings. **d-f)** Histogram of the quadrupole interaction parameters for boron nuclear spins in h-BN with $V_B^-$ defect. $Q_{zz}$ (d), $Q_{xx} + Q_{yy}$ (e), and $Q_{xx} - Q_{yy}$ (f).

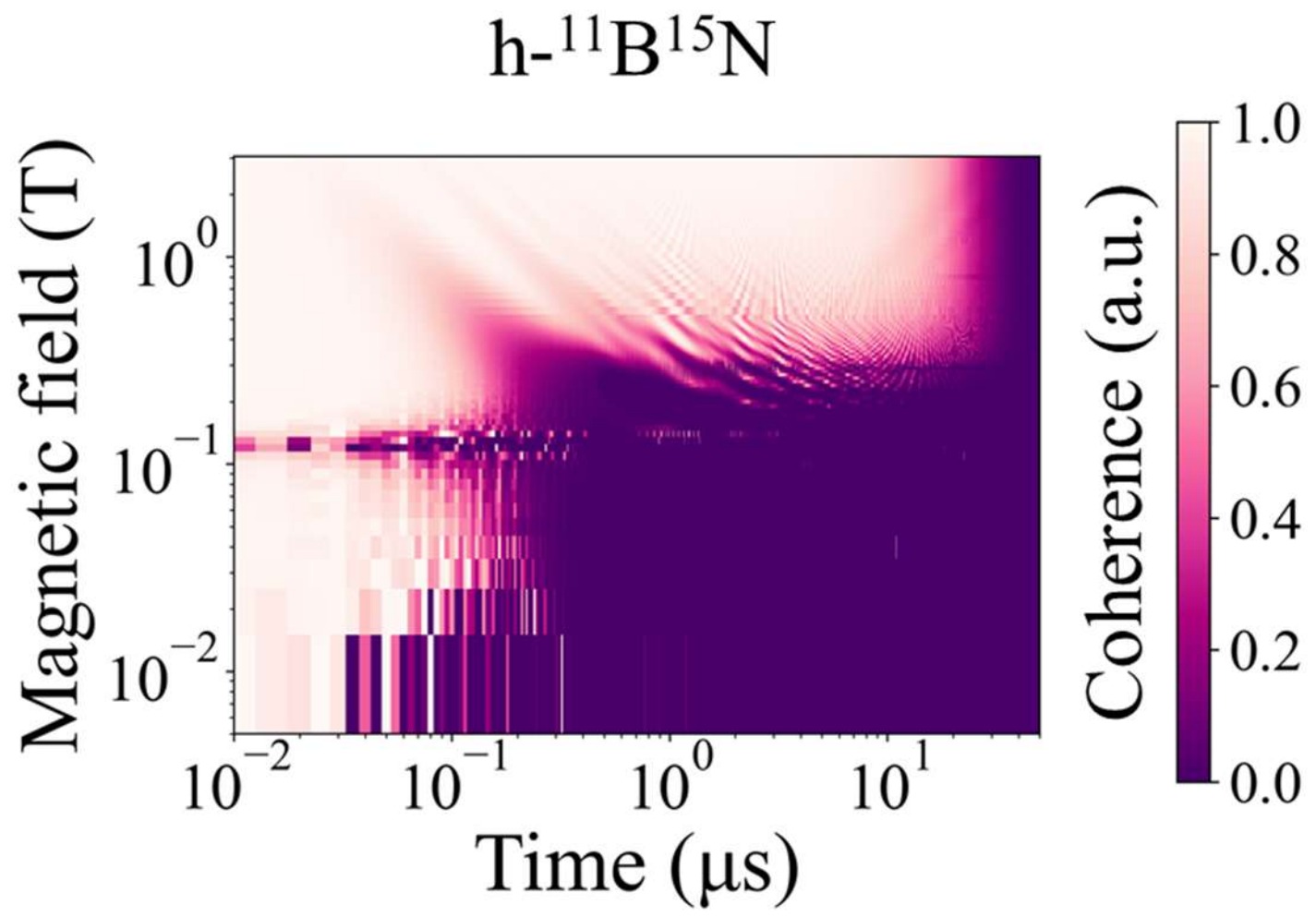

**Figure S8.** Coherence function of the $V_B^-$ spin in h-$^{11}B^{15}N$, computed at the gCCE2 level of theory as a function of external magnetic field strength ranging from 0.01 to 3 T within a time scale up to 50 µs.

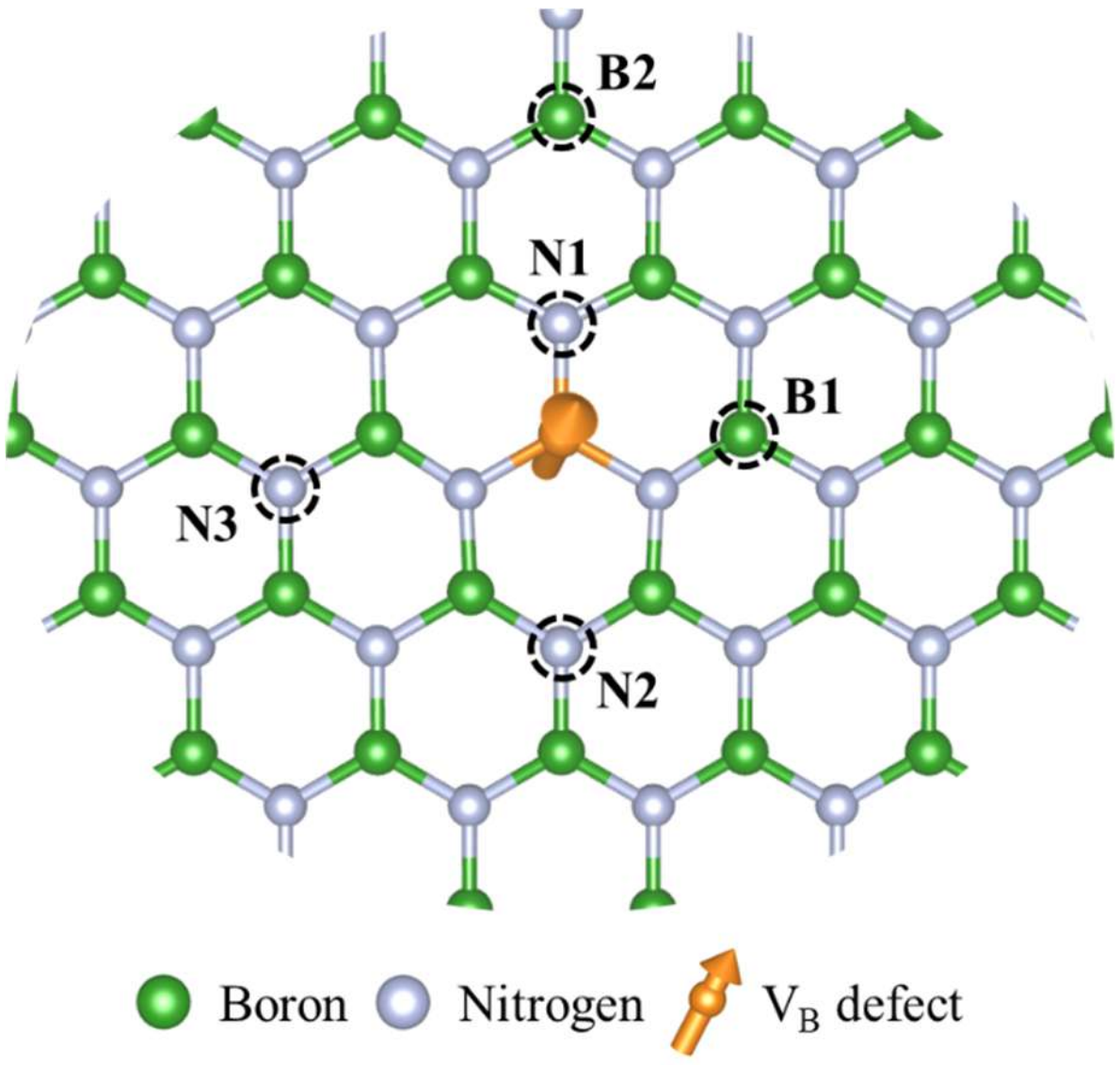

**Figure S9.** Relaxed structure of the $V_B^-$ defect in h-BN. Symmetrically inequivalent atoms around the $V_B^-$ are selected to compare their spin Hamiltonian parameters in Supplementary Table 1. These atoms are labeled as N1, N2, N3, B1, and B2.

**Table S1.** Computed hyperfine parameters for nearby nuclear spins of $V_B^-$ defect in h-BN. The selected sites are shown in Figure S9.

| Atom site | $A_{xx}$ (MHz) | $A_{yy}$ (MHz) | $A_{zz}$ (MHz) | $A_{xy}$ (MHz) |
|---|---|---|---|---|
| $^{14}N1$ | 45.83 | 91.43 | 47.57 | 0 |
| $^{14}N2$ | 0.86 | 0.19 | -0.40 | 0 |
| $^{14}N3$ | 6.73 | 4.49 | 4.35 | 0.61 |
| $^{10}B1$ | 0.84 | 0.18 | -1.56 | 1.17 |
| $^{10}B2$ | -0.41 | 0.04 | -0.58 | 0 |

**Table S2.** Computed quadrupole parameters for nearby nuclear spins of a $V_B^-$ defect in h-BN. The selected sites are shown in Figure S9.

| Atom site | $Q_{xx}$ (MHz) | $Q_{yy}$ (MHz) | $Q_{zz}$ (MHz) | $Q_{xy}$ (MHz) |
|---|---|---|---|---|
| $^{14}$N1 | -0.59 | 1.09 | -0.50 | 0 |
| $^{14}$N2 | 0.01 | -0.01 | 0.00 | 0 |
| $^{14}$N3 | 0.03 | -0.02 | -0.01 | 0.01 |
| $^{10}$B1 | -0.12 | -0.13 | 0.25 | -0.04 |
| $^{10}$B2 | -0.14 | -0.12 | 0.26 | 0 |